\documentclass[letterpaper,twocolumn,11pt,accepted=2025-11-26]{quantumarticle}
\pdfoutput=1
\usepackage[utf8]{inputenc}
\usepackage[english]{babel}
\usepackage[T1]{fontenc}
\usepackage{amsmath}
\usepackage{hyperref}
\usepackage{siunitx}

\usepackage[numbers,sort&compress]{natbib}
\begin{document}

\title{GPU-accelerated Effective Hamiltonian Calculator}

\author[1,2,3,4]{Abhishek Chakraborty}
\orcid{0009-0007-9483-3394}
\email{achakraborty@chapman.edu}
\author[1]{Taylor L. Patti}
\orcid{0000-0002-4242-6072}
\email{tpatti@nvidia.com}
\author[1]{Brucek Khailany}
\orcid{0000-0002-7584-3489}
\author[2,3,4,5]{Andrew N. Jordan}
\orcid{0000-0002-9646-7013}
\author[6]{Anima Anandkumar}
\orcid{0000-0002-6974-6797}
\affil[1]{NVIDIA, Santa Clara, California 95051, USA}
\affil[2]{Center for Coherence and Quantum Science (CCQS), University of Rochester, Rochester, NY 14627, USA}
\affil[3]{Department of Physics and Astronomy, University of Rochester, Rochester, NY 14627, USA}
\affil[4]{Institute for Quantum Studies (IQS), Chapman University, Orange, CA 92866, USA}
\affil[5]{The Kennedy Chair in Physics, Chapman University, Orange, CA 92866, USA}
\affil[6]{Department of Computing + Mathematical Sciences (CMS),
California Institute of Technology (Caltech), Pasadena, CA 91125, USA}
\maketitle

\begin{abstract}
Effective Hamiltonian calculations for large quantum systems can be both analytically intractable and numerically expensive using standard techniques. In this manuscript, we present numerical techniques inspired by Nonperturbative Analytical Diagonalization (NPAD) and the Magnus expansion for the efficient calculation of effective Hamiltonians. While these tools are appropriate for a wide array of applications, we here demonstrate their utility for models that can be realized in circuit-QED settings. Our numerical techniques are available as an open-source Python package, qCH\textsubscript{eff}, which is available on GitHub (\url{https://github.com/NVlabs/qCHeff}) and PyPI (\url{https://pypi.org/project/qcheff/}). We use the CuPy library for GPU-acceleration and report up to 15x speedup on GPU over CPU for NPAD, and up to 42x speedup for the Magnus expansion (compared to QuTiP), for large system sizes.
\end{abstract}

\section{Introduction}

Effective Hamiltonians \cite{Joergensen1975} are an essential tool for analyzing the behavior of complicated quantum systems \cite{Jolicard1995}. Detailed microscopic modeling can quickly become intractable due to the memory and computational costs of representing and operating large Hilbert spaces, which grow exponentially as the number of quantum particles increases. To make such computations tractable, effective models with reduced complexity are used to approximate the properties of interest. Typically, simple truncation fails to account for relevant dynamics, and special care is required when constructing low-energy effective Hamiltonians. Effective Hamiltonians are also useful in reducing the complexity of time-dependent problems \cite{Zeuch2020,  Magesan2020, Consani2020, Venkatraman2022, SandovalSantana2019, Haas2019}. A system with very rapid time-dependence can be approximated by a series of effective time-independent Hamiltonians (a technique known as time coarse-graining) \cite{Macri2023, Gamel2010, Brinkmann2016}, which simplifies the analysis. Effective models have found applications in several fields of physics, including quantum chemistry \cite{Jolicard1995}, condensed matter \cite{Powell2010}, quantum optics \cite{Klimov2002}, quantum information science \cite{James2007, James2000}, and quantum simulation \cite{AntoSztrikacs2023,Bravyi2008,Peng2023a}.

Effective models allow us to analyze a class of physical systems without committing to a specific experimental implementation. Any microscopic physics consistent with the effective interaction is a good candidate. For instance, the examples considered in this article could each correspond to several potential experimental platforms, including: trapped ion systems \cite{James2000}, 2-d superconducting circuits \cite{Blais2004, Blais2007, Greentree2006}, Nitrogen-Vacancy (NV) centers in diamond \cite{Greentree2006}, etc. 

\begin{figure*}[ht!]
    \centering
    \includegraphics[width=\linewidth]{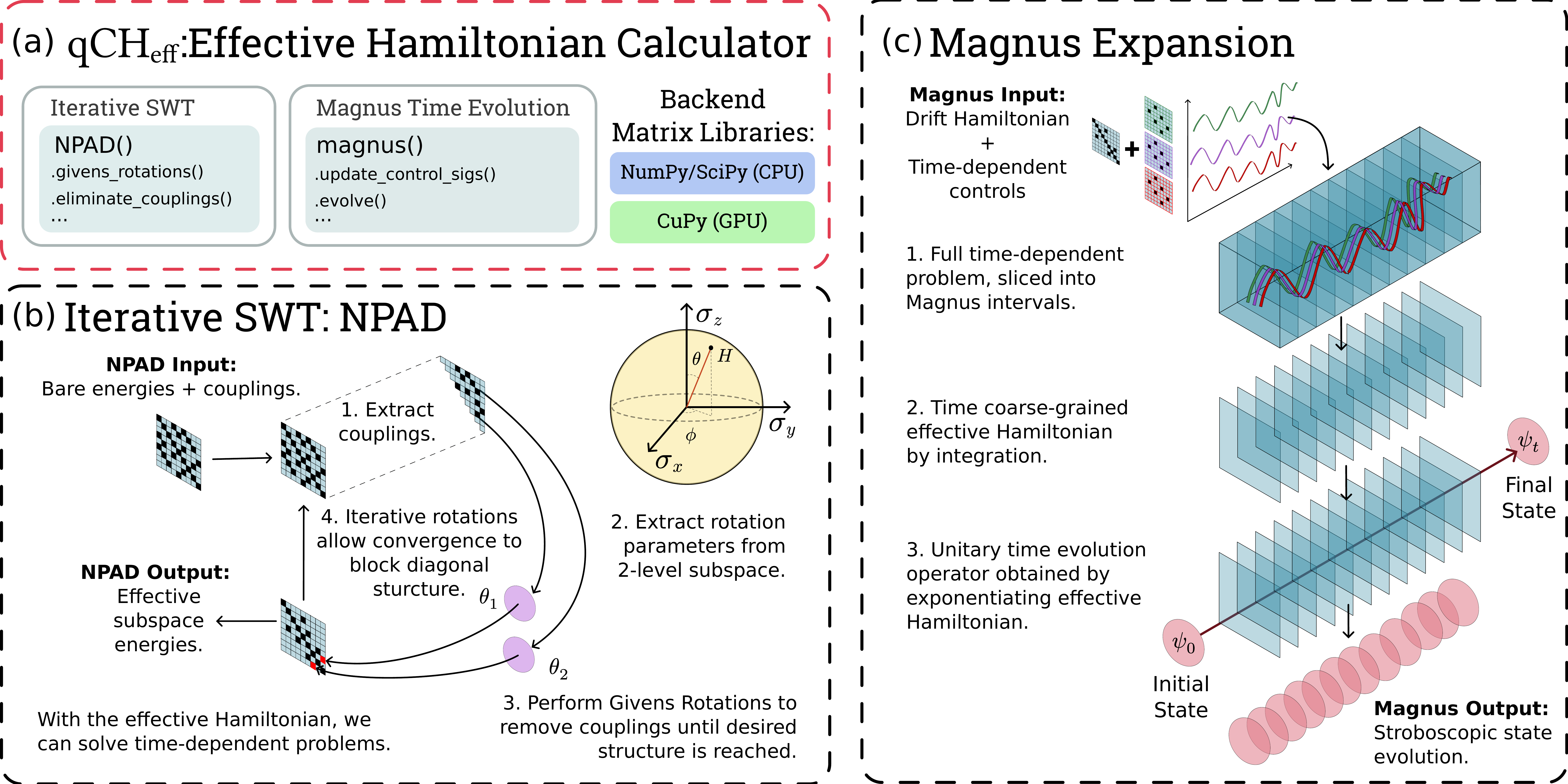}
    \caption{(a) Overview of the qCH\textsubscript{eff} package API. (b) The NPAD algorithm implements an iterative Schrieffer-Wolff transformation. Givens rotations \cite{Givens1958} exactly diagonalize two-level subspaces of the full Hamiltonian and iteration can be stopped early once the desired subspace Hamiltonian/eigenvalues is obtained. (c) The Magnus expansion can accurately and efficiently simulate time-evolution for quantum systems with rapid time dependence.}
    \label{fig:qcheff-schematic}
\end{figure*} 

As an example, let us consider the approximations made when modeling superconducting circuits. The transmon \cite{Koch2007} is the workhorse of the field, however it is not truly a qubit (two-level quantum system), but rather an anharmonic oscillator with many levels. Simply truncating all but the lowest two levels fails to capture the effects of higher energy levels, which should not generally be neglected as they are both an experimental reality and a control scheme tool \cite{Blais2004, Blais2021}. Recent work has demonstrated control of 12 distinct levels in a transmon device \cite{Wang2024}. Readout for these devices is performed using a dispersively coupled resonator, which leads to photon-number dependent frequency shifts of qubit levels \cite{Blais2021}, so these effects must also be captured in such an effective model. 

Several analytical techniques have been developed for effective Hamiltonian calculations, but most such techniques become intractable for all but the simplest systems. For example, the Schrieffer-Wolff transformation (SWT) \cite{Bravyi2011, Li2022} is a perturbative method where an effective block-diagonal Hamiltonian is obtained up to some order in the perturbation parameter. However, the fully-expanded expressions grow exponentially in the number of terms with each increasing order \cite{Day2024}. A numerical approach is thus needed to broaden the range of systems we can analyze. 

To that end, we present qCH\textsubscript{eff}, a Python package with numerical techniques for effective Hamiltonian calculations, including dimensionality reduction and time coarse-graining. For dimensionality reduction, we implement a numerical iterative SWT method based on the Nonperturbative Analytical Diagonalization (NPAD) \cite{Li2022} technique for efficient block diagonalization of high-dimensional Hamiltonians. To our knowledge, this is the first publicly available implementation of this technique. For time coarse-graining, we provide time-evolution methods based on the first-order Magnus expansion \cite{Blanes2009}, thus enabling accurate and efficient simulation for  systems with rapid time-dependence when compared to standard numerical integration of the Schrodinger equation. The techniques in qCH\textsubscript{eff} also provide clear and direct physical insights, something that is usually either inaccessible or requires some computational overhead in commonly used techniques such as numerical diagonalization. Overall, we believe that this package will enable the analysis of very high-dimensional quantum systems with low dimensional effective dynamics and those with rapid time-dependence, beyond what is currently possible. qCH\textsubscript{eff} is available as an open-source Python package on GitHub (\url{https://github.com/NVlabs/qCHeff}) and PyPI (\url{https://pypi.org/project/qcheff/}).

In Section~\ref{sec:qcheff-package}, we give a brief overview of the software package and its design. In Section~\ref{sec:ISWT}, we discuss the iterative Schrieffer-Wolff transformation for obtaining a subset of eigenvalues of a time-independent Hamiltonian. We then apply it to the Jaynes-Cummings-Hubbard (JCH) model in order to examine the Mott-like insulating phase. In Section~\ref{sec:time-coarse-graining}, we discuss average Hamiltonian theory and the Magnus expansion for effective time-independent Hamiltonians for system with rapid time-dependence. We demonstrate that the Magnus expansion is more accurate than the rotating wave approximation, and simulate time-evolution for state transfer in an isotropic spin chain. We also benchmark both methods on both accuracy and running time. For the cases considered in this article, NPAD is up to 15x faster and the Magnus expansion is up to 300x faster on GPU when compared to CPU. The Magnus expansion is also up to 42x faster than direct integrating the Schr\"odinger equation with QuTiP for the same numerical error. Unless specified otherwise, we assume $\hbar=1$ throughout this article.

\section{\texorpdfstring{The qCH\textsubscript{eff} Package}{qCHeff}} \label{sec:qcheff-package}

The qCH\textsubscript{eff} package has two main submodules: Iterative Schrieffer-Wolff transformations (\texttt{qcheff.iswt}) and Magnus time-evolution (\texttt{qcheff.magnus}). 

The Iterative SWT submodule facilitates iterative diagonalization of time-independent Hamiltonians. The main technique is a numerical version of nonperturbative analytical diagonalization (NPAD) \cite{Li2022}. If a given Hamiltonian has known block structure and only a few eigenvalues are required, this iterative technique can prove more efficient than complete numerical diagonalization. The Magnus submodule enables efficient time coarse-grained simulations of Hamiltonians with rapid time-dependence. Simulating such systems using conventional integration techniques can be computationally intensive. 

For both submodules, qCH\textsubscript{eff} provides a high-level interface that dispatches to the appropriate backend device (CPU or GPU), as specified. CPUs are designed for a wide range of tasks and excel at sequential tasks; GPUs can efficiently perform a restricted set of operations in parallel, providing speedups in several numerical tasks. Our numerical implementations take advantage of this speedup on GPU. All methods have been implemented using the NumPy, SciPy (CPU) and CuPy \cite{Okuta2017} (GPU) Python libraries. If the input is sparse, the appropriate sparse matrix library is chosen: \texttt{scipy.sparse} on CPU and \texttt{cupyx.scipy.sparse} on GPU. CuPy is a drop-in replacement for much of the NumPy and SciPy API. Due to the seamlessness of this drop-in design, our Python source code is identical for both CPU/GPU backends. More details about the package are provided in Appendix~\ref{sec:package-appendix}. Further details of the benchmarks are provided in Appendix~\ref{sec:benchmarks}.

\section{Effective Block-Diagonal Hamiltonians}\label{sec:ISWT}

\subsection{Schrieffer-Wolff Transformation}

The Schrieffer-Wolff transformation (SWT) \cite{CohenTannoudji1998} is an operator-level version of degenerate perturbation theory in which a low-energy effective Hamiltonian is obtained from an exact Hamiltonian by a unitary transformation that decouples the low-energy and high-energy subspaces. The SWT has been successfully applied to obtain block-diagonal effective Hamiltonians in some simple cases, such as the cross-resonance gate for transmons \cite{Koch2007,Zhu2013,Magesan2020}. While the SWT is a ubiquitous tool across several fields of physics, the nomenclature varies across fields \cite{Bravyi2011, Li2022}.  

However, this technique is restricted to cases where the energy hierarchy doesn't change appreciably. Additionally, in systems with many interacting quantum modes, spurious resonances between them can make the problem intractable, a common occurrence in the case of superconducting qubit designs with tunable energies, such as the split-junction transmon \cite{Koch2007} and fluxonium \cite{Manucharyan2009} qubits. A review of some other popular methods can be found in Ref. \cite{Paulisch2014}. 

\subsection{Iterative SWT: Numerical NPAD}
\label{sec:npad}
The NPAD method \cite{Li2022} can be understood as a generalization (to Hermitian matrices) of Jacobi iteration \cite{Press2007}, a recursive method for diagonalizing symmetric matrices. As with Jacobi iteration, NPAD will diagonalize the matrix in the limit of infinite iterations. However, if only a few eigenvalues (for example, energies for states belonging to a particular subspace) of the matrix are relevant, we can stop early once the appropriate subspace is decoupled from the rest of the Hilbert space. 

\noindent First, consider the two-level Hamiltonian:
\begin{equation}
\label{eq:tls-ham}
\begin{split}
H =&
\begin{pmatrix}
\epsilon+\delta & ge^{-i\phi}\\
ge^{i\phi} & \epsilon-\delta
\end{pmatrix}\\
&= \epsilon I + \delta \sigma^{z} \\ 
&+ g( \sigma^{x} \cos\phi + \sigma^{y}\sin\phi).
\end{split}
\end{equation}
where $\epsilon$ is a global energy shift, $2\delta$ is the energy splitting between the two levels and $g$ ($\phi$) is the amplitude (phase) of the generally complex-valued coupling between the two levels, and $\sigma^x$, $\sigma^y$ and $\sigma^z$ are the Pauli spin operators.

We can plot this Hamiltonian on a Bloch sphere of radius $\sqrt{g^2 + \delta^2}$, as shown in Fig.~\ref{fig:qcheff-schematic}. The resulting Bloch vector makes an angle $\phi$ with the $x$-axis, and \(\theta=\arctan (g/\delta)\) with the $z$-axis.
Couplings between different eigenstates are represented by off-diagonal elements. 
Hamiltonians can be diagonalized using repeated Givens rotations  \cite{Li2022}, which are  rotations in a two-dimensional subspace of the full Hamiltonian. For example, a Givens rotation about an axis \(\hat{n} = \sigma^{y} \cos\phi - \sigma^{x}\sin\phi\) by an angle \(\theta=\arctan (g/\delta)\) will diagonalize the Hamiltonian in Eq.~\eqref{eq:tls-ham}. Importantly, we can ensure that this rotation does not flip level ordering. Although this involves the inverse tangent, it is possible to replace all trigonometric functions with algebraic expressions for numerical stability, using trigonometric identities. This is the basic iterative step of NPAD. 

For higher-dimensional systems, any two-level subspace in the larger Hilbert space can be diagonalized by an appropriate Givens rotation. Iterating over all the undesired couplings leads to the suppression of those off-diagonal elements of the Hamiltonian and renormalization of the energy levels in proportion to the effect of the coupling. Any Givens rotation will invariably mix other levels, but rotating away the largest off-diagonal terms leads to rapid convergence to an effective block-diagonal Hamiltonian \cite{Li2022}, and will eventually diagonalize the full Hamiltonian. NPAD is more efficient than full diagonalization when we are concerned about only a few select eigenvalues in the spectrum. If all eigenvalues are required, then numerical techniques for diagonalization, such as those based on matrix decomposition, for example, into an an orthogonal matrix and an upper triangular matrix (QR decomposition) or lower and upper triangular matrices (LU decomposition), among others, will be more efficient. For NPAD, we can stop iterating as soon as the desired levels are decoupled from all others. We can also choose the ordering of Givens rotations carefully for faster convergence. If the Hamiltonian already has a block-diagonal structure, we can combine rotations for different blocks together. Additionally, numerical NPAD avoids any ambiguity due to labeling, which is a concern for numerical diagonalization when degeneracies arise.

\subsection{Example: Predicting Mott Lobes in the Jaynes-Cummings-Hubbard model}

\begin{figure}[ht!]
    \centering
    \includegraphics[width=\linewidth]{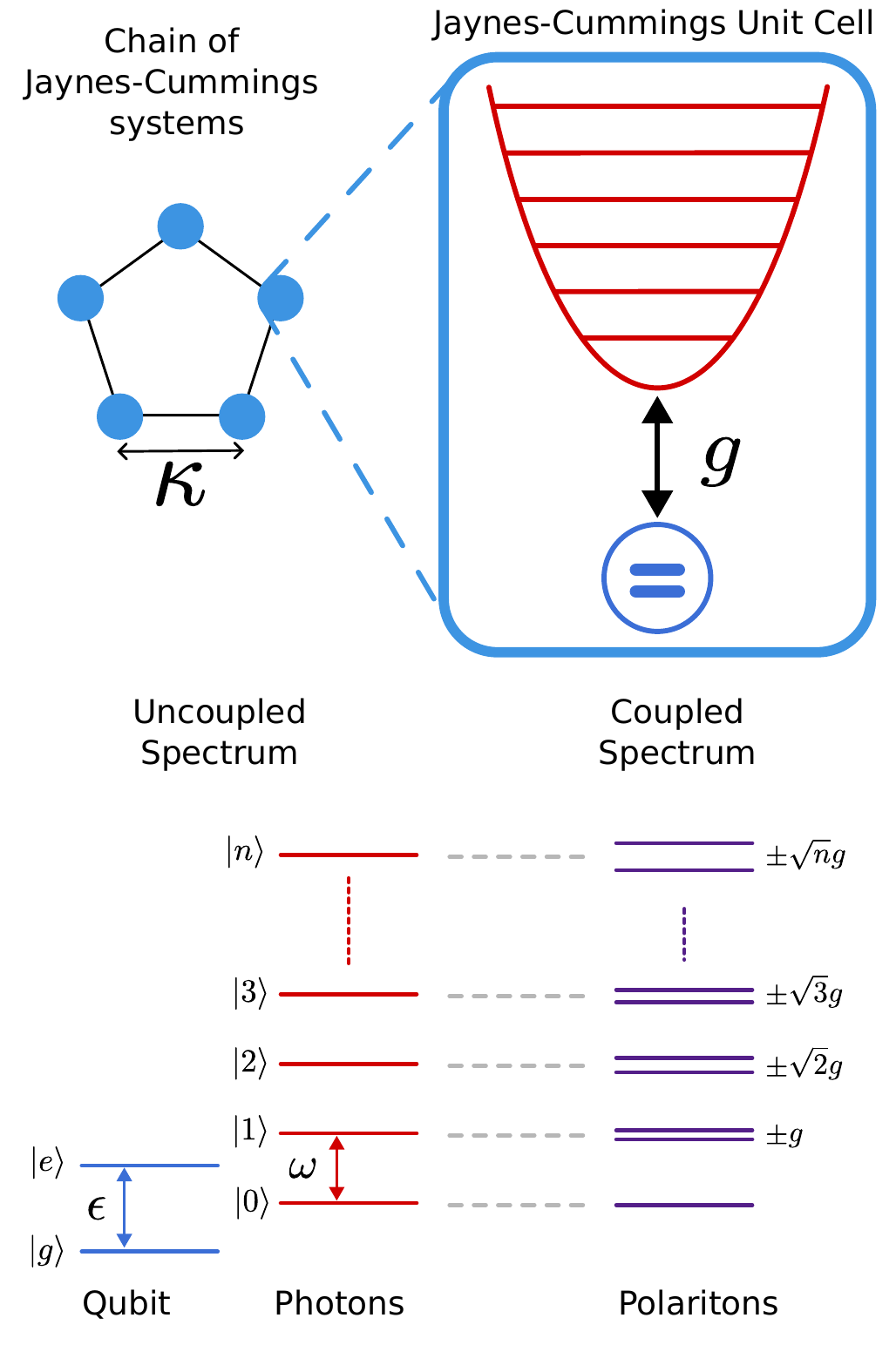}
    \caption{ (Top) Schematic showing the setup for the Jaynes-Cummings-Hubbard (JCH) model, sometimes also called the Jaynes-Cummings (JC) lattice model. Each lattice site (blue circle) has a linear resonator (frequency $\omega$) coupled to an atom (frequency $\epsilon$) with strength $g$. Photons can hop between neighboring resonators at a rate $\kappa$. 
    (Bottom) Uncoupled and coupled JC spectrum.}
    \label{fig:system-jch}
\end{figure}

NPAD is most useful when the bare Hamiltonian has known block structure that we can exploit. To illustrate such a case, we consider the Jaynes-Cummings-Hubbard (JCH) model \cite{Greentree2006, Koch2009}. The JCH model is related to the more widely known Bose-Hubbard model, which describes the many-body physics of a chain of spinless bosons \cite{Fisher1989}.\hfill

The JCH model comprises of a chain of cavities, coupled via a hopping-like interaction between cavity photons, each strongly coupled to a distinct detuned qubit as shown in Fig.~\ref{fig:system-jch}. While this model displays both a Mott-insulator phase and a superfluid state, \cite{Angelakis2007,Greentree2006,Hartmann2008,Koch2009,TwyeffortIrish2008}, we focus on the former for this discussion. While the following discussion is general to all experimental implementations, we note that Nitrogen Vacancy (NV) centers \cite{Greentree2006} and superconducting qubits \cite{Koch2009, Greentree2006} have been proposed as platforms to implement this system. 

\noindent The Hamiltonian for the system is given by:
\begin{equation}
H^{\rm JCH} = \sum_j H^{\rm JC}_j
 -  H^{\rm hop} - \mu N\label{eq:jch-ham}
\end{equation}
The first term is the usual Jaynes-Cummings (JC) Hamiltonian \cite{Jaynes1963}
\begin{equation}
        H^{\rm JC}_j = \omega a_j^\dagger a_j+\epsilon\sigma_j^+\sigma_j^-+g(a_j^\dagger\sigma_j^-+\sigma_j^+a_j),
        \label{eq:jch-jc-term}
\end{equation}
where $a_j^\dagger (a_j)$ is the creation (annihilation) operator for the cavity mode with frequency $\omega$,  $\sigma_j^{\pm}$ are the Pauli spin raising and lowering operators for qubit $j$ with frequency $\epsilon$ and $g$ is the qubit-photon coupling strength. For later convenience, we also define the cavity-qubit detuning $\Delta \equiv \omega - \epsilon$.

The JC model has been studied extensively in quantum optics. The eigenstates are collective qubit-photon excitations called \textit{polaritons} \cite{CohenTannoudji1998}. As shown in Fig.~\ref{fig:system-jch}, the polaritons are organized in parity doublets $|n\pm\rangle$ with energies \cite{Koch2009}
\begin{equation}
E_{n\pm}=n\omega+\Delta/2\pm[(\Delta/2)^2+ng^2]^{1/2},
\label{eq:jc-energies}
\end{equation}
for $n\geq1$ and a disconnected vacuum state $|0\rangle$ with energy $E_0$.

The second term in the JCH Hamiltonian describes the photon-hopping between cavities:
\begin{equation}
H^{\mathrm{hop}}=- \kappa\sum_{\langle i,j\rangle}\left(a_i^\dagger a_j+a_j^\dagger a_i\right),\label{eq:jch-hopping-term}\end{equation}
where $\langle i,j\rangle$ indicates that the sum ranges over nearest neighbors and $\kappa$ is the rate at which photons hop between cavities. The final term in the JCH Hamiltonian comes from the treatment of the model in the grand canonical ensemble. The chemical potential couples directly to the polariton number operator, which is a conserved quantity in the JC Hamiltonian. 

\begin{equation}N= \sum_{j} n_j = \sum_{j}(a_{i}^{\dagger}a_{j}+\sigma_{j}^{+}\sigma_{j}^{-}).\label{eq:jch-polariton-chem-potential-term}\end{equation}
This model shows a phase transition between a Mott insulator-like phase and a superfluid as we tune the ratio $\kappa/g$. In the so-called atomic limit, $\kappa/g\ll 1$, we would expect a Mott phase since the hopping interaction can be treated as a pertubative interaction between neighboring JC unit cell. This phase is characterized by a fixed number of excitations per site with no fluctuations \cite{Greentree2006}. For a small enough $\kappa/g$, the unit cells decouple completely, and the ground state of the full system is a product of the ground state of the onsite Hamiltonians:
\begin{equation}
\begin{split}
    H_j^{\rm(on-site)} &= \omega a_j^\dagger a_j+\varepsilon\sigma_j^+\sigma_j^-+g(a_j^\dagger\sigma_j^-+\sigma_j^+a_j)\\
    & -\mu (a_{j}^{\dagger}a_{j}+\sigma_{j}^{+}\sigma_{j}^{-}),
\end{split}
\end{equation}
which includes a polariton-number ($n$) dependent energy shift due to the chemical potential in addition to the usual JC Hamiltonian. The resulting energies are $E^{\mu}_{(n\pm, 0)} = E_{(n\pm, 0)}- \mu n$, where the site index $j$ has been dropped.

If the chemical potential is very small compared to the characteristic system frequencies, then the ground state is just the vacuum (no polariton) state $|0\rangle$. However, as we tune the quantity $(\omega -\mu)$, admitting a single polariton will be energetically favourable when $E^\mu_0 = E^\mu_{1, -}$. In fact, there are multiple such boundaries separating different Mott lobes, with each neighboring lobe having an extra polariton per site in the ground state. Since the JC model is exactly solvable, an analytic expression can be obtained \cite{Koch2009} for these threshold values:
\begin{equation}\frac{(\mu-\omega)}{g}=\sqrt{n+\left(\frac{\Delta}{2g}\right)^2}-\sqrt{n+1+\left(\frac{\Delta}{2g}\right)^2},\label{eq:mott-lobes-theory}\end{equation}

\noindent where $n$ is the number of polaritons in the ground state in the given qubit-cavity system. We use this expression to check the accuracy of our numerical results. To numerically compute these boundaries, we follow the approach outlined in \cite{Koch2009} for the atomic limit. Since we are interested only in the lowest energy state, the symmetric states $|n+\rangle$ can be neglected, as they have higher energy than the anti-symmetric states. Thus, the Mott lobe boundaries can be computed from the eigenvalues of the anti-symmetric states only.

We compare two different techniques to obtain the first few Mott lobes (except $E^\mu_0 = E^\mu_{1, -}$, for figure clarity, as this boundary differs qualitatively from the rest \cite{Greentree2006}): 
\begin{itemize}
    \item \textbf{Numerical Diagonalization}: We compute the eigenvalues by numerical diagonalization. Then, we extract the relevant eigenvalues using eigenstate-overlap labeling. These operations were performed using routines available in scQubits \cite{Groszkowski2021, Chitta2022}/QuTiP \cite{Johansson2012, Johansson2013}, which are themselves based on SciPy subroutines. 
    \item \textbf{NPAD}: Since the Jaynes-Cummings Hamiltonian has a block structure, with each block having states of the same polariton number, we can rotate away couplings simultaneously. We target all atom-photon couplings below a threshold polariton number (determined by the number of Mott lobes we are interested in). For the plotted lobes, we perform a single unitary transformation using one Givens rotation matrix per lobe, all based on the same initial Hamiltonian.
\end{itemize}
\begin{figure}[t!]
    \centering
    \includegraphics[width=\linewidth]{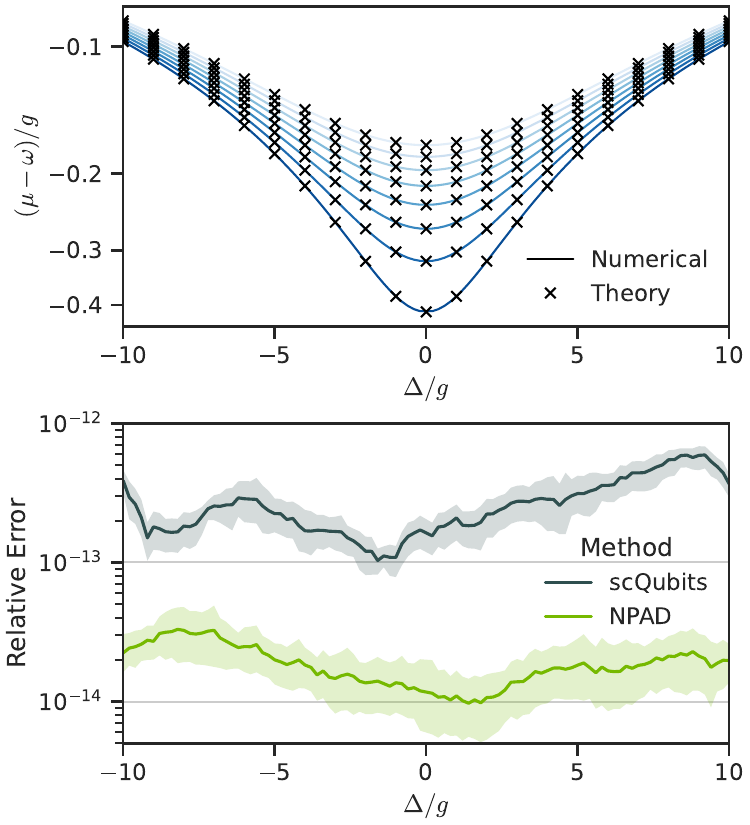}
    \caption{(Top) Chemical potential boundaries between different Mott Lobes for the JCH model in the atomic limit ($\kappa/g\ll 1$), computed using NPAD. Solid lines indicate numerical results and crosses ($\times$) mark values predicted by the analytical expression given by (Eq.~\eqref{eq:mott-lobes-theory}). (Bottom) Relative error for NPAD and numerical diagonalization compared to the exact analytical expression (Eq.~\eqref{eq:mott-lobes-theory}). The solid lines indicate the error averaged over all energy levels/Mott lobes for a given detuning and the shaded region indicates the $95\%$ confidence interval around the mean. In this problem, NPAD has the added benefit of greater numerical precision over exact diagonalization due to fewer numerical operations. NPAD performs only one Givens rotations to calculate each eigenvalue/Mott lobe, which is exact in this case due to the structure and symmetry of the problem. On the other hand, numerical diagonalization solves for the matrix eigenvalues using general diagonalization techniques thus requiring more floating point operations, leading to a higher floating point error. Both methods have error below $10^{-12}$ which does not qualitatively affect predictions in typical scenarios.}
    \label{fig:npad-result} 
\end{figure}
\begin{figure*}[tb!]
    \centering
    \includegraphics[width=\linewidth]{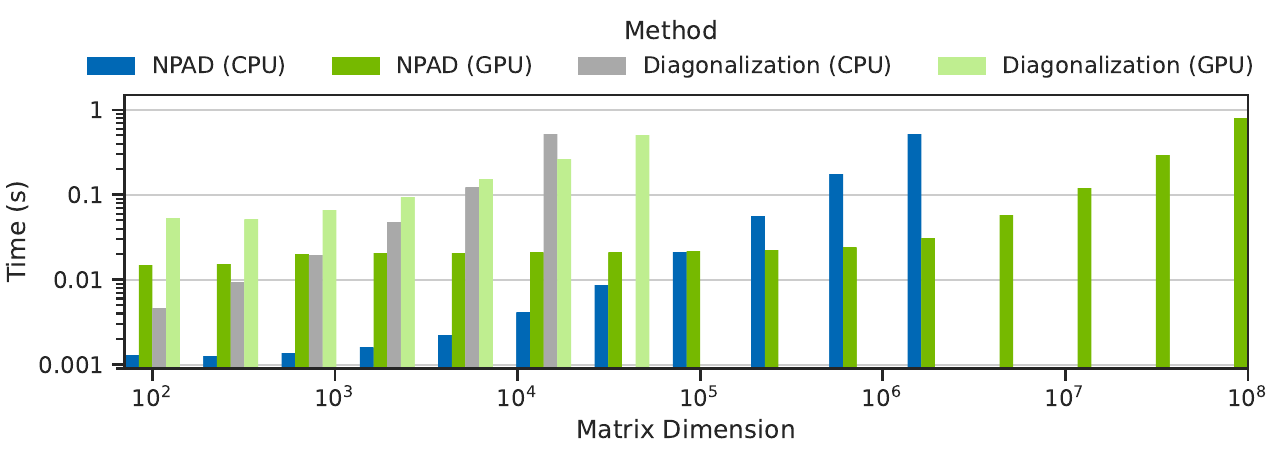}
    \caption{Comparing performance of NPAD and numerical diagonalization on both CPU and GPU. The test sparse Hamiltonian matrix is $H_{\rm test} = a^\dagger a + (a + a^\dagger)$, where $a (a^\dagger)$ is a bosonic mode annihilation (creation) operator truncated to $N$ levels. For NPAD, we benchmark the running time of applying 5 NPAD iterations each with a single Givens rotation to test Hamiltonians of various sizes. We impose a \qty{1}{\second} timeout for each operation on a matrix of a given size and report only parts which finish within this timeout: numerical diagonalization up to matrix dimension around $10^4$ on CPU and $10^5$ on GPU, and NPAD on CPU up to a matrix dimension around $10^6$. On both CPU and GPU, NPAD is roughly an order of magnitude faster compared to numerical diagonalization on the same type of device. The running time for NPAD on GPU for smaller matrices is largely determined by the baseline execution latency on GPUs and remains roughly constant up to a matrix dimension of $10^6$, beyond which the matrix size in memory is comparable to the total available VRAM. A similar plateau is seen on the CPU, but for small matrices. As the matrix size grows in CPU memory, the running time increases rapidly. NPAD is roughly an order of magnitude faster on GPUs, which are optimized for sparse matrix multiplication, compared to CPUs.}
    \label{fig:npad-scaling}
\end{figure*}
The Mott lobe boundaries are shown for the different techniques in Fig.~\ref{fig:npad-result}. Our results are consistent with those obtained in \cite{Koch2009, Greentree2006}. The bottom panel of Fig.~\ref{fig:npad-result} shows the error relative to Eq.~\eqref{eq:mott-lobes-theory} averaged over energy levels/Mott lobes for different values of $\Delta/g$. For this problem, NPAD has the added benefit of greater numerical precision over exact diagonalization due to fewer numerical operations. NPAD performs only one Givens rotations to calculate each eigenvalue/Mott lobe, which is exact in this case due to the structure and symmetry of the problem. On the other hand, numerical diagonalization solves for the matrix eigenvalues using general diagonalization techniques thus requiring more floating point operations, leading to a higher floating point error. The error in both cases is below $10^{-12}$ and QuTiP typically rounds off differences at this order, so we do not expect any qualitative difference between results for the two methods. For example, superconducting qubits typically have frequencies around a few gigahertz. The two methods would only differ by a millihertz, which is well below the resolution of the control electronics, typically $\sim$\qty{1}{\Hz}, used in experiments.

We benchmark running time for NPAD and full numerical diagonalization on both CPU and GPU. We use a sparse test matrix $M_{\rm test} = a^\dagger a + (a + a^\dagger)$, where $a (a^\dagger)$ is a bosonic mode annihilation (creation) operator truncated to $N$ levels. We measure the running time for a typical NPAD operation comprising of 5 consecutive Givens rotations. The choice of couplings is arbitrary and does not affect performance, so we pick the 5 smallest couplings. We also measure the running time for sparse numerical diagonalization routines. Fig.~\ref{fig:npad-scaling} shows results from these benchmarks for sparse matrices with $N$ between $10^2$ and $10^8$. To ensure completion in a reasonable duration, we set a \qty{1}{\second} timeout for given method and matrix size, excluding setup/teardown time and associated memory transfers. We report only those parts that finish within the timeout to ensure a fair comparison. Exact diagonalization is very slow for large matrices, so we only perform this benchmark up to a matrix dimension of $10^4$ on CPU and $10^5$ on GPU. The NPAD technique solves a different problem than standard diagonalization. Standard eigensolvers compute a subset of eigenvalues based on the relative magnitude of eigenvalues. NPAD, instead, calculates specific labeled eigenvalues by considering the effects of a given set of transitions. NPAD is about 15x faster on GPU compared to CPU for large matrices. NPAD on CPU also exceeds the \qty{1}{\second} timeout at about $10^6$ matrix dimension. GPUs are significantly more performant as the matrix size increases. On both CPU and GPU, NPAD is roughly an order of magnitude faster compared to numerical diagonalization on the same type of device. NPAD on GPU for small matrices is mainly limited by the baseline execution latency of GPUs and remains roughly constant up to a matrix dimension of $10^6$, beyond which the matrix size in memory is comparable to the total available VRAM. A similar plateau is seen on the CPU, but for small matrices. As the matrix size grows in CPU memory, the running time increases rapidly. 

\section{Time Coarse-Graining Hamiltonians}\label{sec:time-coarse-graining}

As the leading method for solving the Schr{\"o}dinger equation, time-evolution simulation is among the most crucial tasks in quantum science. Such simulations are used to design control schemes for quantum information processing \cite{Dalgaard2022, Blais2004, Blais2021} and analyzing the behavior of driven systems like those encountered in NMR \cite{Ernst1990}, among other applications. Whether solving the Schr\"odinger equation analytically or numerically, time-dependent Hamiltonians represent a particular challenge, as the analytic solutions are largely unknown if not nonexistent while numerical techniques are computationally intensive \cite{Venkatraman2022, Xiao2022, Machnes2018}.

For example, superconducting qubits are typically controlled using rapidly oscillating microwave pulses \cite{Blais2004, Blais2021, Krantz2019, Rasmussen2021}, Numerical simulations of these control schemes are often used to design quantum gates by optimizing over a large control parameter space, a technique called quantum optimal control \cite{Koch2022, Dalgaard2022, Howard2023, Bondar2023, Chen2023b,Mahesh2023, Machnes2018}. For Hamiltonians with rapid time dependence, the requirement of a small integration step size \cite{Press2007} can be a performance bottleneck in the optimization process. Furthermore, the integration cannot be performed in parallel for different time steps, as the state at any given time step depends on that of the previous time step. Hence, it is challenging to simulate deep circuits such as those used in randomized benchmarking or quantum algorithms, at the pulse level.

The above limitations of integration motivate the development of more efficient simulation methods for time-dependent Hamiltonians. A variety of such techniques have been developed. For example, systems with rapid time-dependence are usually analyzed in a rotating frame, which is equivalent to an interaction picture treatment \cite{Sakurai2020}. While there are several possible choices of such a rotating frame, for driven quantum systems, it is common to choose a frame rotating at the frequency of the drive. The rotating wave approximation (RWA) \cite{CohenTannoudji1998} drops rapidly oscillating terms corresponding to far-detuned transitions in this frame, for which the oscillations are assumed to average to zero on the scale of a larger integration time step. In simple cases, this approximation can provide a time-independent Hamiltonian, but it fails to do so for the vast majority of systems, as this technique is prone to errors in even the simplest cases where the zero integration assumption fails. For example, the RWA leads to an apparent shift, called the Bloch-Siegert shift \cite{Bloch1940} (see Sec.~\ref{sec:magnus-rwa}), in the frequency of a strongly driven qubit. Incorrect application of the RWA can also lead to acausal dynamics \cite{Milonni1995}.
    
 For cases where the RWA is a poor approximation, more sophisticated techniques have been developed. An adaptive RWA technique \cite{Baker2018} was used to study the long-time asymptotic behaviour of open quantum systems. Time-dependent Schrieffer-Wolff perturbation theory has been used \cite{Petrescu2023} to analyze parametric two-qubit gates for coupled transmons under strong driving conditions. Floquet methods \cite{Mundada2020, Forney2010, Son2009} have been useful for cases of periodic drives. Several time coarse-graining methods have also been developed \cite{Gamel2010,SandovalSantana2019,James2007,Joergensen1975,Venkatraman2022,Shirley1965,Zeuch2020a,Xiao2022,Theis2017,Malekakhlagh2020a,Petrescu2020,Malekakhlagh2020,Shillito2021,Zeuch2020,Haas2019,Macri2023,Puzzuoli2023,Petrescu2023}. One such method, the Magnus expansion \cite{Marecek2020,Sanchez2011,Butcher2009,Kopylov2019,Dalgaard2022,Geiser2014,Bamber2020,Arnal2021,Vogl2019}, was selected for our software package over other techniques due to its simple yet accurate modeling.

 \subsection{Magnus Expansion}
Average Hamiltonian theory \cite{Brinkmann2016} has recently gained popularity in quantum information settings \cite{Puzzuoli2023, Macri2023, Dalgaard2022}. In this approach, a time-dependent Hamiltonian $H(t)$ is approximated over a finite time interval by an effective time-independent Hamiltonian $\overline{H}$, which results in the same time-evolution over that period as the original Hamiltonian. The time-independent effective Hamiltonian is usually expressed as a Magnus expansion \cite{Blanes2009} 
 \begin{equation}
 \overline{H} =  \overline{H}^{(1)} + \overline{H}^{(2)} + \overline{H}^{(3)}+ ...,
     \label{eq:magnus-exp}
 \end{equation}
 where the term $\overline{H}^{(n)}$ corresponds to the $n$th order averaged Hamiltonian. In contrast to the Dyson expansion \cite{Dyson1949, Sakurai2020}, the Magnus expansion is Hermitian at every order, such that the corresponding propagator is guaranteed to be unitary. The Magnus expansion for a generally time-dependent Hamiltonian \(H(t)\) can be written as \(H(t) = \sum\nolimits_{k=1}^{\infty} \overline{H}^{(k)} \), where 
 \begin{equation}
\label{eq:magnus-avg}
\begin{split}
    \overline{H}^{(1)} = & \int_0^t H\left(t_1\right) d t_1, \\
    \overline{H}^{(2)} = & -\frac{i}{2} \int_0^t d t_1 \int_0^{t_1} d t_2\left[H\left(t_1\right), H\left(t_2\right)\right],
\end{split}
\end{equation}
\noindent and higher orders continue this pattern of nested commutators.

It is common to truncate the series at low orders because higher order terms involve nested commutators, which are computationally expensive to evaluate. We stress that, even at the lowest order, the Magnus expansion is more accurate than the RWA \cite{Macri2023}. Whereas the RWA accumulates errors at all times by neglecting rapidly oscillating terms, the Magnus expansion remains stroboscopically exact at multiples of the coarse-graining period \(\Delta t\).  Since the Magnus expansion effectively averages the rapidly oscillating terms over a Magnus interval, the time-evolution as calculated using the Magnus expansion agrees exactly with the full Hamiltonian in the rotating frame at only these exact points in time. Since we average the time-evolution over each Magnus interval, the technique cannot make predictions for the time-evolution within each interval.

 While a large part of average Hamiltonian theory literature focuses on analytical uses of the technique, the Magnus expansion has recently \cite{Dalgaard2022, Koch2022} been applied to quantum optimal control problems as an alternative to numerically integrating the Schr\"odinger equation. For numerical simulation, the Magnus expansion allows us to integrate more accurately and using larger time steps than RWA, as its errors remain relatively well controlled. However, there is a trade-off between accuracy and performance. For the same control pulse with a fixed gate duration, using more Magnus intervals will lead to higher accuracy at the cost of longer running time. Additionally, since the propagator for each Magnus interval is guaranteed to be unitary, techniques like quantum process tomography can be used analyze the effect of the applied controls to the system \cite{Blais2021}.

The qCH\textsubscript{eff} package implements first-order Magnus expansion-based time-evolution. The general procedure is shown in Fig.~\ref{fig:qcheff-schematic}. Following the approach in \cite{Dalgaard2022}. we assume the system Hamiltonian is given by 
\begin{equation}
    H(t) = H_0 + \sum_k u^{(k)}(t) H_k, 
    \label{eq:magnus-generic-ham}
\end{equation}
where $H_0$ is the drift Hamiltonian and $u^{(k)}(t)$ is the control pulse for time-independent control Hamiltonian $H_k$. Most systems we are interested in can be written in this form. From Eq.~\eqref{eq:magnus-avg}, we can write down the first order Magnus term for the $n$th interval
\begin{equation}
\overline{H}^{(1)}_n = \Delta t H_0 + \sum_k c^{(1)}_{k,n} H_k.
    \label{eq:magnus-first-order-term}
\end{equation}
The coefficients 
\begin{equation}
c^{(1)}_{k,n} = \int_{t_n}^{t_n + \Delta t} u^{(k)}(t_1) d t_1,
    \label{eq:magnus-first-order-coeff}
\end{equation}
can be computed using the control signals alone and then the effective time-independent Hamiltonian is computed Eq.~\eqref{eq:magnus-first-order-term} and exponentiated to obtain the propagator over the Magnus interval of duration $\Delta t$.
\subsection{Example: Strongly driven qubit: Comparison with RWA}
\begin{figure}[ht!]
    \centering
    \includegraphics[width=0.9\linewidth]{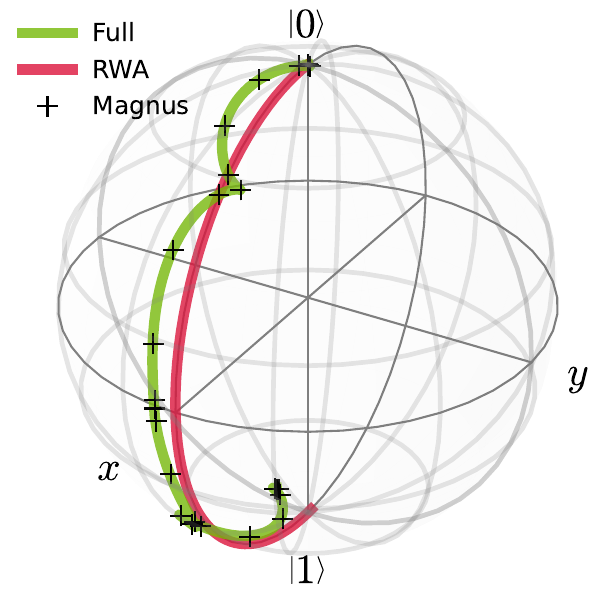}
    \caption{Comparing the Magnus expansion to the RWA for a strongly driven qubit with drive strength is $\Omega_0/\omega = 0.33$, $\omega$ being the qubit frequency. On resonance, the RWA is no longer a good approximation, however the Magnus expansion predicts the correct stroboscopic time-evolution. The full simulation is done with 5000 time steps while the Magnus time-evolution uses just 50 Magnus intervals.}
    \label{fig:magnus-rwa}
\end{figure}

We compare the accuracy of the Magnus expansion to the RWA for the simple case of a driven qubit, with results shown in Fig.~\ref{fig:magnus-rwa}. This system was previously studied using an analytical variant of the Magnus expansion employing a Taylor series approximation for control pulses \cite{Zeuch2020}, where the RWA was shown to fail in the case of strong driving. The Hamiltonian for such a system in the lab frame is given by
\begin{equation}
    H_{\rm lab} = \frac{\omega}{2}\sigma^{z}+\frac{\Omega(t)}{2}\cos(\omega t)\sigma^{x},
    \label{eq:driven-spin-lab-frame}
\end{equation}
where $\omega$ is both the qubit and drive frequency (on resonance), and $\Omega(t)$ is the drive envelope with maximum amplitude $\Omega_0$.
We then transform to a frame rotating with frequency $\omega$, resulting in the following Hamiltonian.
\begin{equation}
H_{\rm rot}(t) = \frac{\Omega(t)}{4}(\sigma^{x}+\cos(2\omega t)\sigma^{x} - \sin(2\omega t)\sigma^{y}).\label{eq:magnus-rot-ham}\end{equation}
The rotating wave approximation (RWA) drops the last two terms, giving us
\begin{equation}
H_{\rm RWA}(t) = \frac{\Omega(t)}{4}\sigma^{x}.\label{eq:magnus-rwa-ham}\end{equation}
The RWA Hamiltonian varies less rapidly compared to Eq.~\eqref{eq:magnus-rot-ham}, but fails to account for Stark-shifted qubit frequency (the Bloch-Siegert shift \cite{Bloch1940}) due to the drive. To make a fair comparison, we simulate both Hamiltonians in Eq.~\eqref{eq:magnus-rot-ham} and Eq.~\eqref{eq:magnus-rwa-ham} using 5000 Magnus intervals. This gives us the ``true'' state evolution for both cases. Next, we simulate time-evolution using only 50 Magnus intervals. Here, we consider $\Omega_0/\omega = 0.33$, where the RWA fails catastrophically, as seen in Fig.~\ref{fig:magnus-rwa}. With only 50 intervals, the stroboscopic Magnus time-evolution remains close to the true time-evolution.  

\subsection{Example: Simulating state transfer in a degenerate spin chain.}\label{sec:magnus-rwa}

To illustrate the effectiveness of the Magnus expansion for solving time-dependent problems, we consider a spin chain with periodic boundary conditions, Ising-type nearest-neighbor and next-nearest-neighbor interactions, described by the drift Hamiltonian
\begin{equation}
    \hat{H}_d = \frac{1}{2} \sum_j \omega_j \sigma^z_j - J\sum_j \sigma^z_j \sigma^z_{j+1} - g\sum_j \sigma^z_j \sigma^z_{j+2}.
    \label{eq:magnus_drift_ham}
\end{equation}

This system has been analyzed in the context of quantum optimal control \cite{Dalgaard2022}. For a degenerate spin chain ($\omega_j = \omega$), we want to find an optimal pulse that effects state transfer between the degenerate ground states $|00...00\rangle\equiv|\mathbf{0}\rangle$ and $|11...11\rangle\equiv|\mathbf{1}\rangle$ using other excited states. To control the spin chain, we apply two global control fields coupling to the Pauli $X$ and $Y$ operators. 
\begin{equation}
    \hat{H}_c (t) = \sum_{k\in\{x, y\}} u^k(t) \sum_j \sigma^k_j.
    \label{eq:magnus_control_ham}
\end{equation}

\begin{figure}[t!]
    \centering
        \includegraphics[width=\linewidth]{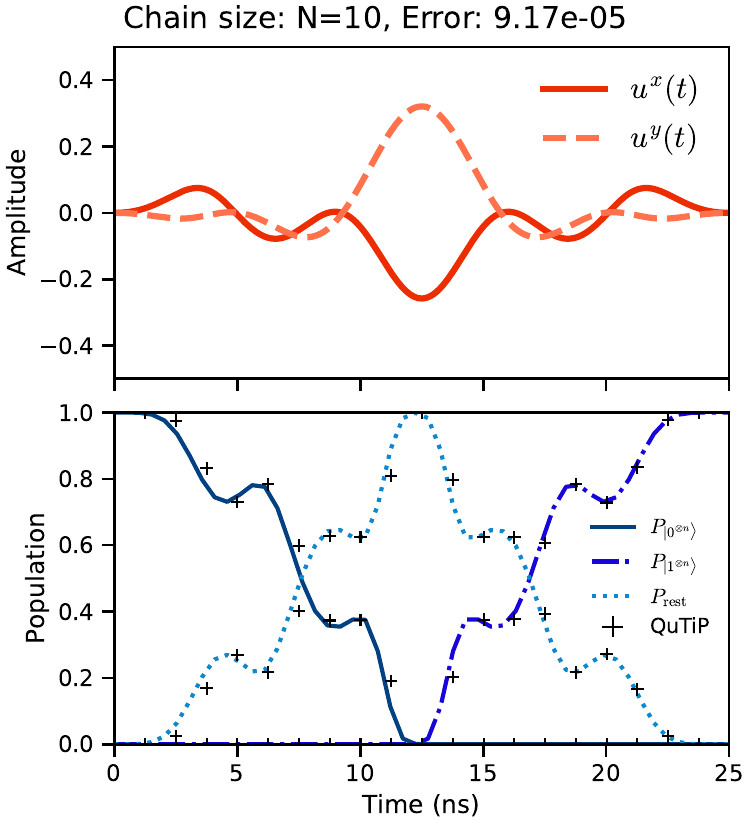}
    \caption{(Top) Pulse shapes for optimal controls to effect population transfer in a degenerate spin chain and (bottom) corresponding population dynamics calculated using the Magnus expansion with 50 Magnus intervals (solid lines) and full numerical integration using QuTiP with 1000 time points (markers) for a chain with 10 spins. For visual clarity, we only measure the two degenerate ground state populations. All other state populations are lumped together into $P_{\rm rest}$.}
    \label{fig:magnus-example}
\end{figure}

\noindent The total Hamiltonian is then given by 
\begin{equation}
    {H} = {H}_d + {H}_c (t).
    \label{eq:magnus_total_ham}
\end{equation}
To frame this as an optimal control problem, we will try to find control pulses $u^x(t)$ and $u^y(t)$ that effect the desired state transfer. Starting from the $|\mathbf{0}\rangle$ state, we evolve the system under the action of the controls to obtain the final state $|\psi(t)\rangle$. We can then minimize the infidelity of the state transfer
\begin{equation}
    \text{Error} = 1 - |\langle\mathbf{1}|\psi(t)\rangle|^2.
\end{equation}
We numerically simulate the time-evolution for the total Hamiltonian given by Eq.~\eqref{eq:magnus_total_ham} using the Magnus expansion for one such optimal pulse shown in Fig.~\ref{fig:magnus-example}. This pulse results in a final state error  of about $10^{-4}$. 

\begin{figure}[tb!]
     \centering
     \includegraphics[width=\linewidth]{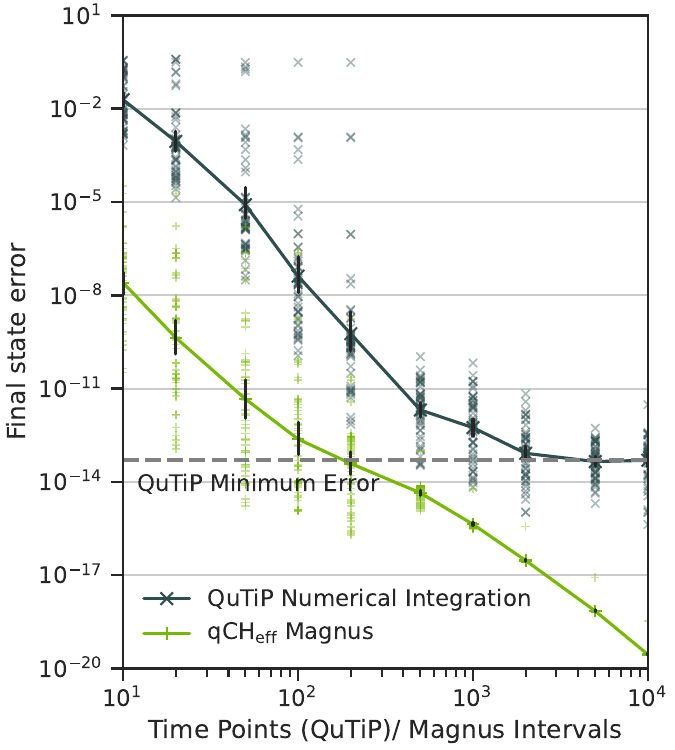}
     \caption{Error in the final predicted state for varying number of integration time steps/Magnus intervals. The markers show the outcome of each individual run. For each method and a given number of time steps/Magnus intervals, the solid line indicating the error averaged over 50 simulations with the control pulse envelope modulated sinusoidally at a random frequency uniformly distributed between $0$ and \qty{10}{\GHz}, a typical parameter regime for superconducting qubits. Vertical lines indicate the 95\% confidence interval around the mean. The variance in final state error decreases for both methods as we increase the number of integration point/Magnus intervals. We see that the Magnus expansion based time evolution achieves an error of $10^{-13}$ with only 200 Magnus intervals, whereas the QuTiP simulation needs 10000 steps for the same accuracy, as Magnus averages over the rapid time dependence while retaining the effects on the system time-evolution.}
     \label{fig:magnus-err}
 \end{figure}
To demonstrate the advantages of the Magnus subroutine, we simulate time evolution of $|\mathbf{0}\rangle$ state for a given pulse. We evaluate the subroutine on two different criteria: accuracy and runtime, and compare it to QuTiP's exact numerical integration of the Schr\"odinger equation. As before, we simulate the time-evolution given a rapidly varying drive for the system using a large number of time steps (QuTiP) or Magnus intervals (Magnus) to get the ``true'' final state. We then evolve the system for a fewer time steps/Magnus intervals and compare the resulting final state to the true final state in each case.  It is important to only compare states from the same method, since the nature of numerical errors might differ between the methods. However, we have confirmed that both QuTiP and Magnus predict approximately the same true final state, with the difference being inconsequential to our state error calculation for either method. The results are shown in Fig.~\ref{fig:magnus-err}. The markers show the outcome of each individual simulation run. For each method and a given number of time steps/Magnus intervals, the solid line indicating the error averaged over 50 simulations with the control pulse envelope modulated sinusoidally at a random frequency uniformly distributed between $0$ and \qty{10}{\GHz}, a typical parameter regime for superconducting qubits. Vertical lines indicate the 95\% confidence interval around the mean. The mean value and variance of final state error decreases for both methods as we increase the number of integration points/Magnus intervals. Whereas QuTiP needs 10000 steps for an error of $10^{-13}$, Magnus need only 200 intervals for the same accuracy by averaging over the rapid time dependence while retaining the effects on the system time-evolution. 

\begin{figure}[t!]
    \centering
    \includegraphics[width=\linewidth]{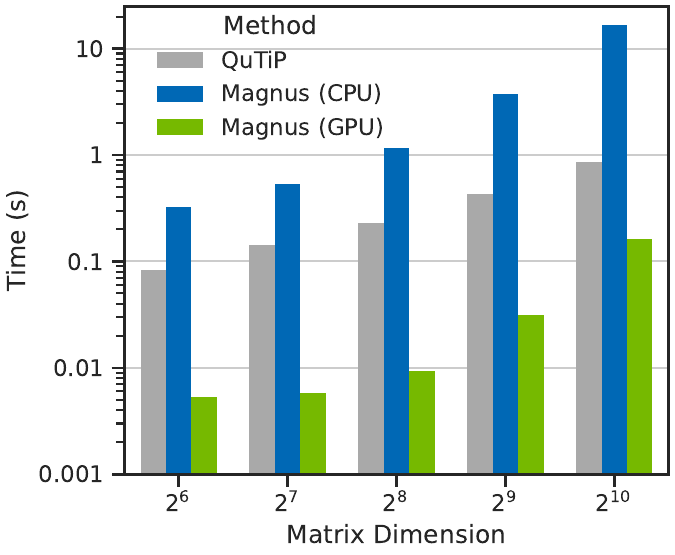}
    \caption{Comparing time-evolution runtime for QuTiP and Magnus, excluding setup and teardown time and memory transfer overheads. We compare simulations that lead to the same final state error relative to the ``true" final state for each respective method. Simulations are for a spin chain of length 6-10 for a \qty{25}{\ns} control pulse. The QuTiP simulation uses 10000 time steps, whereas the same error can be achieved with only 200 Magnus intervals. In this comparison, Magnus is up to 305x faster on GPU when compared to CPU. Additionally, Magnus on GPU is up to 42x faster compared to QuTiP's numerical integration for this setup. Both CPU and GPU implementations of Magnus show a roughly quadratic trend in running time. Magnus on CPU can be slower than numerical integration due to the large memory footprint of the problem. CPUs are not optimized for massive parallelization and can be very slow in performing the \texttt{expm()} operation. On the GPU, embarrasingly parallel execution of \texttt{expm()} over different Magnus intervals ameliorates this problem. The simulations for large matrix sizes nearly saturate the VRAM, also contributing to the quadratic trend. Further optimization of the GPU matrix exponentiation scheme would provide additional speedup.}
    \label{fig:magnus-scaling}
\end{figure}
Next, we compare the running time for both techniques for parameters that lead to the same final state error value of approximately $10^{-13}$. For QuTiP, this requires 10000 time steps. However, we can achieve the accuracy with only 200 Magnus intervals. Results are shown in Fig.~\ref{fig:magnus-scaling}. In the cases we considered, Magnus on GPU is up to 42x faster than integration-based time-evolution using QuTiP. That being said, the memory-footprint of each method is variable, depending on the problem size and specifications. Numerical integration schemes, like those used by QuTiP, are inherently sequential: the steps must be carried out in sequence and cannot be done in parallel. Therefore, at any given time, the memory requirement is determined only by the size of the system Hamiltonian. 
In our implementation, the Magnus expansion is computed for all Magnus intervals in parallel. Naturally, the memory cost increases with the number of Magnus intervals. However, computation can be much faster on GPUs since operations can be parallelized massively, especially at larger scales. Our implementation is up to 305x faster running on the GPU compared to CPU. Fig.~\ref{fig:magnus-scaling} shows this behavior. The Magnus expansion is significantly more efficient, in terms of the accuracy-performance tradeoff, when compared to standard numerical integration. 
Magnus on CPU can be slower than numerical integration for large matrix sizes as a result of the large memory footprint of this technique. Time-evolution for a given state vector is implemented using the matrix exponential \texttt{expm()}. Matrix exponentiation is computationally intensive and CPUs are not optimized for such operations. In the GPU implementation, this is ameliorated by the speedup from automatic embarrassing parallelization of the \texttt{expm()} over different Magnus intervals. Both CPU and GPU implementations of Magnus show a roughly quadratic trend in running time. We attribute this behavior to the memory allocation behavior of \texttt{expm()}. The larger matrix sizes in our benchmark are comparable to the total CPU/GPU memory, contributing to this trend. 
Further optimization of the GPU implementation of \texttt{expm()} could achieve even better scaling, but that is beyond the scope of this work.

\section{Conclusion}
Our Python package, qCH\textsubscript{eff}, implements numerical techniques for effective Hamiltonian calculations. For time-independent problems where only a fraction of the total eigenvalues are required, an iterative SWT technique like NPAD can be more efficient than full-blown numerical diagonalization, because we can strategically select Givens rotations such that few iterations are required. This technique is especially useful when the system has a block-diagonal like structure. Givens rotations are roughly 15x faster on GPU using the CuPy library. To the best of our knowledge, ours is the first publicly available implementations of NPAD.

For problems involving rapid time-dependence, the Magnus expansion can be used to obtain effective time-independent Hamiltonians which are equivalent to the effect of the original Hamiltonian over one Magnus interval. Another advantage over conventional numerical integration techniques is that the Magnus expansion for the Hamiltonian is Hermitian at every order. Even at the lowest order, the Magnus expansion is far more accurate than the popular RWA method, which breaks down for strong drives. We have shown how the Magnus expansion-based approach is more favourable in both accuracy and runtime when compared to QuTiP's numerical integration subroutines with rapidly-varying control signals. Magnus time-evolution is up to 300x faster on GPU compared to CPU. Magnus (GPU) is also up to 42x faster than exact numerical integration on the CPU for the same final state error. Furthermore, the Magnus expansion returns the effective time-independent Hamiltonian and the corresponding propagator for each interval, allowing us to directly interpret the action of our control drives.

Other techniques for effective Hamiltonian calculations may be added to the package in the future, including higher-order Magnus expansions and techniques to handle time-dependent Hamiltonians using iterative SWTs. An analytical extension of NPAD to time-dependent system, called recursive DRAG, has been recently used to suppress coherent errors for the cross-resonance gate in transmons \cite{Li2023}. Other matrix library backends may also be added: automatic differentiation (implemented using libraries like Pytorch or JAX) can enable use of qCH\textsubscript{eff} alongside GPU-accelerated optimization, or deep learning. A sparse implementation of Magnus on GPU would enable simulation of even larger quantum systems. Comparisons to other sparse GPU-accelerated time-evolution methods would also help highlight the performance and utility of our method. However, at the time of writing, we are unaware of any GPU-accelerated solvers or \texttt{expm()} implementations that works for general sparse matrices. The Magnus-based perturbative solver in Qiskit Dynamics \cite{Puzzuoli2023}, inspired by \texttt{DySolve} \cite{Shillito2021} computes the Magnus expansion using the Dyson series with an additional pre-compilation step, making it difficult to compare accuracy and performance to our direct Magnus implementation. 

We believe that qCH\textsubscript{eff} will enable analysis of large-scale systems that were hitherto intractable, while also providing clear physical insight that is erased during other numerical diagonalization procedures. For example, a system consisting of a fluxonium qubit \cite{Manucharyan2009} dispersively coupled to a cavity was examined in \cite{Zhu2013} using perturbation theory. The NPAD technique could be used to scale up such an analysis to multiple coupled qubits, which could be used to design tunable couplers \cite{Moskalenko2021, Weiss2022}. Additionally, crosstalk between different qubits on the same chip (for semi/super-conducting qubits) significantly limits our ability to scale up coherence of large-scale quantum computers. As a result of crosstalk, there are spurious multi-qubit interactions that degrade the quality of control and readout operations. Some mitigation techniques have been developed for correction of spurious two-qubit interactions \cite{Weiss2022, Ding2023, Rosenfeld2024}. A high dimensional Magnus-based simulation of a full chip-scale quantum circuit could shed light on possible mitigation schemes for larger systems. For many-body physics problems like the JCH model, numerical time-evolution has been restricted to systems with only a few cavities, due to the growing size of the Hilbert space \cite{Angelakis2007, TwyeffortIrish2008}. Smaller effective Hamiltonians combined with Magnus-based time-evolution could allow analysis of larger chains and investigate long-range correlation effects, with applications to design of quantum simulators and quantum storage devices. 

\textit{Note added} - While preparing this manuscript, we became aware of recent theory work demonstrating linear scaling for the complexity of computing higher-order perturbative corrections in the SWT \cite{Day2024}.

\section{Acknowledgments}
We thank Justin Dressel, Rayleigh Parker, Habtamu Walelign, Felix Motzoi and Boxi Li for enlightening discussions. We also appreciate the continued support of James Kelly from the Chapman University IS\&T Research Computing Support team. A.C. was supported in part by NVIDIA Corporation and the Army Research Office through Grant no. W911NF-22-1-0258. 

\bibliography{qcheff_paper}

\begin{thebibliography}{10}

\bibitem{Joergensen1975}
Flemming J{\o}rgensen.
\newblock ``Effective hamiltonians''.
\newblock \href{https://dx.doi.org/10.1080/00268977500100971}{Molecular Physics {\bf 29}, 1137--1164}~(1975).

\bibitem{Jolicard1995}
Georges Jolicard.
\newblock ``Effective {Hamiltonian Theory} and {Molecular Dynamics}''.
\newblock \href{https://dx.doi.org/10.1146/annurev.pc.46.100195.000503}{Annual Review of Physical Chemistry {\bf 46}, 83--108}~(1995).

\bibitem{Zeuch2020}
Daniel Zeuch, Fabian Hassler, Jesse~J. Slim, and David~P. DiVincenzo.
\newblock ``Exact rotating wave approximation''.
\newblock \href{https://dx.doi.org/10.1016/j.aop.2020.168327}{Annals of Physics {\bf 423}, 168327}~(2020).

\bibitem{Magesan2020}
Easwar Magesan and Jay~M. Gambetta.
\newblock ``Effective {Hamiltonian} models of the cross-resonance gate''.
\newblock \href{https://dx.doi.org/10.1103/PhysRevA.101.052308}{Physical Review A {\bf 101}, 052308}~(2020).

\bibitem{Consani2020}
Gioele Consani and Paul~A Warburton.
\newblock ``Effective {{Hamiltonians}} for interacting superconducting qubits: Local basis reduction and the {{Schrieffer}}--{{Wolff}} transformation''.
\newblock \href{https://dx.doi.org/10.1088/1367-2630/ab83d1}{New Journal of Physics {\bf 22}, 053040}~(2020).

\bibitem{Venkatraman2022}
Jayameenakshi Venkatraman, Xu~Xiao, Rodrigo~G. Corti{\~{n}}as, Alec Eickbusch, and Michel~H. Devoret.
\newblock ``Static {Effective Hamiltonian} of a {Rapidly Driven Nonlinear System}''.
\newblock \href{https://dx.doi.org/10.1103/PhysRevLett.129.100601}{Physical Review Letters {\bf 129}, 100601}~(2022).
\newblock  \href{http://arxiv.org/abs/2108.02861}{arxiv:2108.02861}.

\bibitem{SandovalSantana2019}
Juan~Carlos Sandoval-Santana, Victor~Guadalupe Ibarra-Sierra, Jos{\'{e}}~Luis Cardoso, Alejandro Kunold, Pedro Roman-Taboada, and Gerardo Naumis.
\newblock ``Method for {Finding} the {Exact Effective Hamiltonian} of {Time-Driven Quantum Systems}''.
\newblock \href{https://dx.doi.org/10.1002/andp.201900035}{Annalen der Physik {\bf 531}, 1900035}~(2019).

\bibitem{Haas2019}
Holger Haas, Daniel Puzzuoli, Feihao Zhang, and David~G. Cory.
\newblock ``Engineering effective {Hamiltonians}''.
\newblock \href{https://dx.doi.org/10.1088/1367-2630/ab4525}{New Journal of Physics {\bf 21}, 103011}~(2019).

\bibitem{Macri2023}
Nicola Macr{\`{i}}, Luigi Giannelli, Elisabetta Paladino, and Giuseppe Falci.
\newblock ``Coarse-{Grained Effective Hamiltonian} via the {Magnus Expansion} for a {Three-Level System}''.
\newblock \href{https://dx.doi.org/10.3390/e25020234}{Entropy {\bf 25}, 234}~(2023).

\bibitem{Gamel2010}
Omar Gamel and Daniel F.~V. James.
\newblock ``Time-averaged quantum dynamics and the validity of the effective {Hamiltonian} model''.
\newblock \href{https://dx.doi.org/10.1103/PhysRevA.82.052106}{Physical Review A {\bf 82}, 052106}~(2010).

\bibitem{Brinkmann2016}
Andreas Brinkmann.
\newblock ``Introduction to average {Hamiltonian} theory. {I}. {Basics}''.
\newblock \href{https://dx.doi.org/10.1002/cmr.a.21414}{Concepts in Magnetic Resonance Part A {\bf 45A}, e21414}~(2016).

\bibitem{Powell2010}
Ben~J. Powell.
\newblock ``Introduction to effective low-energy hamiltonians in condensed matter physics and chemistry''.
\newblock \href{https://dx.doi.org/10.1002/9780470930779.ch10}{Chapter~10, pages 309--366}.
\newblock John Wiley \& Sons, Ltd. ~(2011).
\newblock  \href{http://arxiv.org/abs/0906.1640}{arXiv:0906.1640}.

\bibitem{Klimov2002}
A.~B. Klimov, L.~L. S{\'{a}}nchez-Soto, A.~Navarro, and E.~C. Yustas.
\newblock ``Effective {Hamiltonians} in quantum optics: a systematic approach''.
\newblock \href{https://dx.doi.org/10.1080/09500340210134675}{Journal of Modern Optics {\bf 49}, 2211--2226}~(2002).

\bibitem{James2007}
D~F James and J~Jerke.
\newblock ``Effective {Hamiltonian} theory and its applications in quantum information''.
\newblock \href{https://dx.doi.org/10.1139/p07-060}{Canadian Journal of Physics {\bf 85}, 625--632}~(2007).

\bibitem{James2000}
D.~F.~V. James.
\newblock ``Quantum {Computation} with {Hot} and {Cold Ions}: {An Assessment} of {Proposed Schemes}''.
\newblock \href{https://dx.doi.org/10.1002/3527603182.ch5}{Scalable Quantum Computers: Paving the Way to Realization {\bf 48}, 53--67}~(2000).

\bibitem{AntoSztrikacs2023}
Nicholas Anto-Sztrikacs, Ahsan Nazir, and Dvira Segal.
\newblock ``Effective-{Hamiltonian Theory} of {Open Quantum Systems} at {Strong Coupling}''.
\newblock \href{https://dx.doi.org/10.1103/PRXQuantum.4.020307}{PRX Quantum {\bf 4}, 020307}~(2023).
\newblock  \href{http://arxiv.org/abs/2211.05701}{arxiv:2211.05701}.

\bibitem{Bravyi2008}
Sergey Bravyi, David~P. DiVincenzo, Daniel Loss, and Barbara~M. Terhal.
\newblock ``Simulation of {Many-Body Hamiltonians} using {Perturbation Theory} with {Bounded-Strength Interactions}''.
\newblock \href{https://dx.doi.org/10.1103/PhysRevLett.101.070503}{Physical Review Letters {\bf 101}, 070503}~(2008).
\newblock  \href{http://arxiv.org/abs/0803.2686}{arxiv:0803.2686}.

\bibitem{Peng2023a}
Bo~Peng, Yuan Su, Daniel Claudino, Karol Kowalski, Guang Hao~Low, and Martin Roetteler.
\newblock ``Quantum {Simulation} of {Boson-Related Hamiltonians}: {Techniques}, {Effective Hamiltonian Construction}, and {Error Analysis}''.
\newblock \href{https://dx.doi.org/10.1088/2058-9565/adbf42}{Quantum Science and Technology {\bf 10}, 023002}~(2025).
\newblock  \href{http://arxiv.org/abs/2307.06580}{arXiv:2307.06580}.

\bibitem{Blais2004}
Alexandre Blais, Ren-Shou Huang, Andreas Wallraff, S.~M. Girvin, and R.~J. Schoelkopf.
\newblock ``Cavity quantum electrodynamics for superconducting electrical circuits: {{An}} architecture for quantum computation''.
\newblock \href{https://dx.doi.org/10.1103/PhysRevA.69.062320}{Physical Review A {\bf 69}, 062320}~(2004).

\bibitem{Blais2007}
Alexandre Blais, Jay Gambetta, A.~Wallraff, D.~I. Schuster, S.~M. Girvin, M.~H. Devoret, and R.~J. Schoelkopf.
\newblock ``Quantum-information processing with circuit quantum electrodynamics''.
\newblock \href{https://dx.doi.org/10.1103/PhysRevA.75.032329}{Physical Review A {\bf 75}, 032329}~(2007).

\bibitem{Greentree2006}
Andrew~D. Greentree, Charles Tahan, Jared~H. Cole, and Lloyd C.~L. Hollenberg.
\newblock ``Quantum phase transitions of light''.
\newblock \href{https://dx.doi.org/10.1038/nphys466}{Nature Physics {\bf 2}, 856--861}~(2006).

\bibitem{Givens1958}
Wallace Givens.
\newblock ``Computation of {Plain Unitary Rotations Transforming} a {General Matrix} to {Triangular Form}''.
\newblock \href{https://dx.doi.org/10.1137/0106004}{Journal of the Society for Industrial and Applied Mathematics {\bf 6}, 26--50}~(1958).

\bibitem{Koch2007}
Jens Koch, Terri~M. Yu, Jay Gambetta, A.~A. Houck, D.~I. Schuster, J.~Majer, Alexandre Blais, M.~H. Devoret, S.~M. Girvin, and R.~J. Schoelkopf.
\newblock ``Charge-insensitive qubit design derived from the {Cooper} pair box''.
\newblock \href{https://dx.doi.org/10.1103/PhysRevA.76.042319}{Physical Review A {\bf 76}, 042319}~(2007).

\bibitem{Blais2021}
Alexandre Blais, Arne~L. Grimsmo, S.~M. Girvin, and Andreas Wallraff.
\newblock ``Circuit {{Quantum Electrodynamics}}''.
\newblock \href{https://dx.doi.org/10.1103/RevModPhys.93.025005}{Reviews of Modern Physics {\bf 93}, 025005}~(2021).
\newblock  \href{http://arxiv.org/abs/2005.12667}{arxiv:2005.12667}.

\bibitem{Wang2024}
Zihao Wang, Rayleigh~W. Parker, Elizabeth Champion, and Machiel~S. Blok.
\newblock ``Systematic study of {High} {${E}_J/{E}_C$} transmon qudits up to {$d = 12$}''.
\newblock \href{https://dx.doi.org/10.1103/physrevapplied.23.034046}{Physical Review Applied {\bf 23}, 034046}~(2025).
\newblock  \href{http://arxiv.org/abs/2407.17407}{arXiv:2407.17407}.

\bibitem{Bravyi2011}
Sergey Bravyi, David~P. DiVincenzo, and Daniel Loss.
\newblock ``Schrieffer--{Wolff} transformation for quantum many-body systems''.
\newblock \href{https://dx.doi.org/10.1016/j.aop.2011.06.004}{Annals of Physics {\bf 326}, 2793--2826}~(2011).

\bibitem{Li2022}
Boxi Li, Tommaso Calarco, and Felix Motzoi.
\newblock ``Nonperturbative {Analytical Diagonalization} of {Hamiltonians} with {Application} to {Circuit QED}''.
\newblock \href{https://dx.doi.org/10.1103/PRXQuantum.3.030313}{PRX Quantum {\bf 3}, 030313}~(2022).

\bibitem{Day2024}
Isidora Araya~Day, Sebastian Miles, Hugo Kerstens, Daniel Varjas, and Anton~R. Akhmerov.
\newblock ``Pymablock: An algorithm and a package for quasi-degenerate perturbation theory''.
\newblock \href{https://dx.doi.org/10.21468/scipostphyscodeb.50}{SciPost Physics Codebases}~(2025).
\newblock  \href{http://arxiv.org/abs/2404.03728}{arXiv:2404.03728}.
\newblock  code:~\href{10.21468/SciPostPhysCodeb.50-r2.1}{10.21468/SciPostPhysCodeb.50-r2.1}.

\bibitem{Blanes2009}
S.~Blanes, F.~Casas, J.~A. Oteo, and J.~Ros.
\newblock ``The {Magnus} expansion and some of its applications''.
\newblock \href{https://dx.doi.org/10.1016/j.physrep.2008.11.001}{Physics Reports {\bf 470}, 151--238}~(2009).

\bibitem{Okuta2017}
Ryosuke Okuta, Yuya Unno, Daisuke Nishino, Shohei Hido, and Crissman Loomis.
\newblock ``Cupy: A numpy-compatible library for nvidia gpu calculations''.
\newblock In Proceedings of Workshop on Machine Learning Systems (LearningSys) in The Thirty-first Annual Conference on Neural Information Processing Systems (NIPS).
\newblock ~(2017).
\newblock  url:~\url{http://learningsys.org/nips17/assets/papers/paper_16.pdf}.
\newblock  code:~\href{https://github.com/cupy/cupy}{cupy/cupy}.

\bibitem{CohenTannoudji1998}
Claude Cohen-Tannoudji, Jacques Dupont-Roc, and Gilbert Grynberg.
\newblock ``Atom-{Photon Interactions}: {Basic Processes} and {Applications}''.
\newblock \href{https://dx.doi.org/10.1002/9783527617197}{John Wiley \& Sons}. ~(1998).

\bibitem{Zhu2013}
Guanyu Zhu, David~G. Ferguson, Vladimir~E. Manucharyan, and Jens Koch.
\newblock ``Circuit {QED} with fluxonium qubits: {Theory} of the dispersive regime''.
\newblock \href{https://dx.doi.org/10.1103/PhysRevB.87.024510}{Physical Review B {\bf 87}, 024510}~(2013).
\newblock  \href{http://arxiv.org/abs/1210.1605}{arXiv:1210.1605}.

\bibitem{Manucharyan2009}
Vladimir~E. Manucharyan, Jens Koch, Leonid~I. Glazman, and Michel~H. Devoret.
\newblock ``Fluxonium: Single cooper-pair circuit free of charge offsets''.
\newblock \href{https://dx.doi.org/10.1126/science.1175552}{Science {\bf 326}, 113--116}~(2009).

\bibitem{Paulisch2014}
Vanessa Paulisch, Han Rui, Hui~Khoon Ng, and Berthold-Georg Englert.
\newblock ``Beyond adiabatic elimination: {A} hierarchy of approximations for multi-photon processes''.
\newblock \href{https://dx.doi.org/10.1140/epjp/i2014-14012-8}{The European Physical Journal Plus {\bf 129}, 12}~(2014).

\bibitem{Press2007}
William~H. Press.
\newblock ``Numerical {Recipes} 3rd {Edition}: {The Art} of {Scientific Computing}''.
\newblock Cambridge University Press. Cambridge [u.a.]~(2007).
\newblock 3. ed. edition.
\newblock  url:~\url{https://numerical.recipes/}.

\bibitem{Koch2009}
Jens Koch and Karyn Le~Hur.
\newblock ``Superfluid--{Mott}-insulator transition of light in the {Jaynes-Cummings} lattice''.
\newblock \href{https://dx.doi.org/10.1103/PhysRevA.80.023811}{Physical Review A {\bf 80}, 023811}~(2009).

\bibitem{Fisher1989}
Matthew P.~A. Fisher, Peter~B. Weichman, G.~Grinstein, and Daniel~S. Fisher.
\newblock ``Boson localization and the superfluid-insulator transition''.
\newblock \href{https://dx.doi.org/10.1103/PhysRevB.40.546}{Physical Review B {\bf 40}, 546--570}~(1989).

\bibitem{Angelakis2007}
Dimitris~G. Angelakis, Marcelo~Franca Santos, and Sougato Bose.
\newblock ``Photon-blockade-induced {Mott} transitions and {$XY$} spin models in coupled cavity arrays''.
\newblock \href{https://dx.doi.org/10.1103/PhysRevA.76.031805}{Physical Review A {\bf 76}, 031805}~(2007).

\bibitem{Hartmann2008}
M.J. Hartmann, F.G.S.L. Brandão, and M.B. Plenio.
\newblock ``Quantum many-body phenomena in coupled cavity arrays''.
\newblock \href{https://dx.doi.org/10.1002/lpor.200810046}{Laser \& Photonics Reviews {\bf 2}, 527--556}~(2008).
\newblock  \href{http://arxiv.org/abs/0808.2557}{arXiv:0808.2557}.

\bibitem{TwyeffortIrish2008}
E.~K. Twyeffort~Irish, C.~D. Ogden, and M.~S. Kim.
\newblock ``Polaritonic characteristics of insulator and superfluid states in a coupled-cavity array''.
\newblock \href{https://dx.doi.org/10.1103/PhysRevA.77.033801}{Physical Review A {\bf 77}, 033801}~(2008).

\bibitem{Jaynes1963}
E.T. Jaynes and F.W. Cummings.
\newblock ``Comparison of quantum and semiclassical radiation theories with application to the beam maser''.
\newblock \href{https://dx.doi.org/10.1109/PROC.1963.1664}{Proceedings of the IEEE {\bf 51}, 89--109}~(1963).

\bibitem{Groszkowski2021}
Peter Groszkowski and Jens Koch.
\newblock ``Scqubits: a {Python} package for superconducting qubits''.
\newblock \href{https://dx.doi.org/10.22331/q-2021-11-17-583}{Quantum {\bf 5}, 583}~(2021).

\bibitem{Chitta2022}
Sai~Pavan Chitta, Tianpu Zhao, Ziwen Huang, Ian Mondragon-Shem, and Jens Koch.
\newblock ``Computer-aided quantization and numerical analysis of superconducting circuits''.
\newblock \href{https://dx.doi.org/10.1088/1367-2630/ac94f2}{New Journal of Physics {\bf 24}, 103020}~(2022).
\newblock  \href{http://arxiv.org/abs/2206.08320}{arXiv:2206.08320}.

\bibitem{Johansson2012}
J.~R. Johansson, P.~D. Nation, and Franco Nori.
\newblock ``{QuTiP}: {An} open-source {Python} framework for the dynamics of open quantum systems''.
\newblock \href{https://dx.doi.org/10.1016/j.cpc.2012.02.021}{Computer Physics Communications {\bf 183}, 1760--1772}~(2012).

\bibitem{Johansson2013}
J.~R. Johansson, P.~D. Nation, and Franco Nori.
\newblock ``{QuTiP} 2: {A Python} framework for the dynamics of open quantum systems''.
\newblock \href{https://dx.doi.org/10.1016/j.cpc.2012.11.019}{Computer Physics Communications {\bf 184}, 1234--1240}~(2013).

\bibitem{Dalgaard2022}
Mogens Dalgaard and Felix Motzoi.
\newblock ``Fast, high precision dynamics in quantum optimal control theory''.
\newblock \href{https://dx.doi.org/10.1088/1361-6455/ac6366}{Journal of Physics B: Atomic, Molecular and Optical Physics {\bf 55}, 085501}~(2022).
\newblock  \href{http://arxiv.org/abs/2110.06187}{arxiv:2110.06187}.

\bibitem{Ernst1990}
Richard~R. Ernst, Geoffrey Bodenhausen, and {and}~Alexander Wokaun.
\newblock ``Principles of {Nuclear Magnetic Resonance} in {One} and {Two Dimensions}''.
\newblock \href{https://dx.doi.org/10.1093/oso/9780198556473.001.0001}{International {Series} of {Monographs} on {Chemistry}}. Oxford University Press. Oxford, New York~(1990).

\bibitem{Xiao2022}
Z.~Xiao, E.~Doucet, T.~Noh, L.~Ranzani, R.W. Simmonds, L.C.G. Govia, and A.~Kamal.
\newblock ``Perturbative {Diagonalization} for {Time-Dependent Strong Interactions}''.
\newblock \href{https://dx.doi.org/10.1103/PhysRevApplied.18.024009}{Physical Review Applied {\bf 18}, 024009}~(2022).
\newblock  \href{http://arxiv.org/abs/2103.09260}{arxiv:2103.09260}.

\bibitem{Machnes2018}
Shai Machnes, Elie Ass\'emat, David Tannor, and Frank~K. Wilhelm.
\newblock ``Tunable, flexible, and efficient optimization of control pulses for practical qubits''.
\newblock \href{https://dx.doi.org/10.1103/PhysRevLett.120.150401}{Physical Review Letters {\bf 120}, 150401}~(2018).
\newblock  \href{http://arxiv.org/abs/1507.04261v2}{arXiv:1507.04261v2}.

\bibitem{Krantz2019}
Philip Krantz, Morten Kjaergaard, Fei Yan, Terry~P. Orlando, Simon Gustavsson, and William~D. Oliver.
\newblock ``A {{Quantum Engineer}}'s {{Guide}} to {{Superconducting Qubits}}''.
\newblock \href{https://dx.doi.org/10.1063/1.5089550}{Applied Physics Reviews {\bf 6}, 021318}~(2019).
\newblock  \href{http://arxiv.org/abs/1904.06560}{arxiv:1904.06560}.

\bibitem{Rasmussen2021}
S.E. Rasmussen, K.S. Christensen, S.P. Pedersen, L.B. Kristensen, T.~B{\ae}kkegaard, N.J.S. Loft, and N.T. Zinner.
\newblock ``Superconducting {{Circuit Companion}}---an {{Introduction}} with {{Worked Examples}}''.
\newblock \href{https://dx.doi.org/10.1103/PRXQuantum.2.040204}{PRX Quantum {\bf 2}, 040204}~(2021).

\bibitem{Koch2022}
Christiane~P. Koch, Ugo Boscain, Tommaso Calarco, Gunther Dirr, Stefan Filipp, Steffen~J. Glaser, Ronnie Kosloff, Simone Montangero, Thomas Schulte-Herbr{\"{u}}ggen, Dominique Sugny, and Frank~K. Wilhelm.
\newblock ``Quantum optimal control in quantum technologies. {Strategic} report on current status, visions and goals for research in {Europe}''.
\newblock \href{https://dx.doi.org/10.1140/epjqt/s40507-022-00138-x}{EPJ Quantum Technology {\bf 9}, 1--60}~(2022).

\bibitem{Howard2023}
Joel Howard, Alexander Lidiak, Casey Jameson, Bora Basyildiz, Kyle Clark, Tongyu Zhao, Mustafa Bal, Junling Long, David~P. Pappas, Meenakshi Singh, and Zhexuan Gong.
\newblock ``Implementing two-qubit gates at the quantum speed limit''.
\newblock \href{https://dx.doi.org/10.1103/PhysRevResearch.5.043194}{Physical Review Research {\bf 5}, 043194}~(2023).

\bibitem{Bondar2023}
Denys~I. Bondar, Llorenç~Balada Gaggioli, Georgios Korpas, Jakub Marecek, Jiri Vala, and Kurt Jacobs.
\newblock ``Globally optimal control of quantum dynamics''.
\newblock \href{https://dx.doi.org/10.1103/g4fb-xm13}{Physical Review Research{\bf 7}}~(2025).
\newblock  \href{http://arxiv.org/abs/2209.05790}{arXiv:2209.05790}.

\bibitem{Chen2023b}
Yuquan Chen, Yajie Hao, Ze~Wu, Bi-Ying Wang, Ran Liu, Yanjun Hou, Jiangyu Cui, Man-Hong Yung, and Xinhua Peng.
\newblock ``Accelerating quantum optimal control through iterative gradient-ascent pulse engineering''.
\newblock \href{https://dx.doi.org/10.1103/PhysRevA.108.052603}{Physical Review A {\bf 108}, 052603}~(2023).

\bibitem{Mahesh2023}
T.~S. Mahesh, Priya Batra, and M.~Harshanth Ram.
\newblock ``Quantum optimal control: practical aspects and diverse methods.''.
\newblock \href{https://dx.doi.org/10.1007/s41745-022-00311-2}{J. Indian Inst. Sci. {\bf 103}, 591--607}~(2023).

\bibitem{Sakurai2020}
J.~J. Sakurai and Jim Napolitano.
\newblock ``Modern quantum mechanics''.
\newblock \href{https://dx.doi.org/10.1017/9781108587280}{Cambridge University Press}. Cambridge~(2020).
\newblock Third edition.

\bibitem{Bloch1940}
F.~Bloch and A.~Siegert.
\newblock ``Magnetic {Resonance} for {Nonrotating Fields}''.
\newblock \href{https://dx.doi.org/10.1103/PhysRev.57.522}{Physical Review {\bf 57}, 522--527}~(1940).

\bibitem{Milonni1995}
P.~W. Milonni, D.~F.~V. James, and H.~Fearn.
\newblock ``Photodetection and causality in quantum optics''.
\newblock \href{https://dx.doi.org/10.1103/PhysRevA.52.1525}{Physical Review A {\bf 52}, 1525--1537}~(1995).

\bibitem{Baker2018}
Brian Baker, Andy C.~Y. Li, Nicholas Irons, Nathan Earnest, and Jens Koch.
\newblock ``Adaptive rotating-wave approximation for driven open quantum systems''.
\newblock \href{https://dx.doi.org/10.1103/PhysRevA.98.052111}{Physical Review A {\bf 98}, 052111}~(2018).

\bibitem{Petrescu2023}
Alexandru Petrescu, Camille~Le Calonnec, Catherine Leroux, Agustin Di~Paolo, Pranav Mundada, Sara Sussman, Andrei Vrajitoarea, Andrew~A. Houck, and Alexandre Blais.
\newblock ``Accurate methods for the analysis of strong-drive effects in parametric gates''.
\newblock \href{https://dx.doi.org/10.1103/PhysRevApplied.19.044003}{Physical Review Applied {\bf 19}, 044003}~(2023).
\newblock  \href{http://arxiv.org/abs/2107.02343}{arxiv:2107.02343}.

\bibitem{Mundada2020}
Pranav~S. Mundada, Andr{\'{a}}s Gyenis, Ziwen Huang, Jens Koch, and Andrew~A. Houck.
\newblock ``Floquet-{{Engineered Enhancement}} of {{Coherence Times}} in a {{Driven Fluxonium Qubit}}''.
\newblock \href{https://dx.doi.org/10.1103/PhysRevApplied.14.054033}{Physical Review Applied {\bf 14}, 054033}~(2020).

\bibitem{Forney2010}
Anne~M. Forney, Steven~R. Jackson, and Frederick~W. Strauch.
\newblock ``Multi-frequency control pulses for multi-level superconducting quantum circuits''.
\newblock \href{https://dx.doi.org/10.1103/PhysRevA.81.012306}{Physical Review A {\bf 81}, 012306}~(2010).
\newblock  \href{http://arxiv.org/abs/0909.1577}{arxiv:0909.1577}.

\bibitem{Son2009}
Sang-Kil Son, Siyuan Han, and Shih-I Chu.
\newblock ``Floquet formulation for the investigation of multiphoton quantum interference in a superconducting qubit driven by a strong ac field''.
\newblock \href{https://dx.doi.org/10.1103/PhysRevA.79.032301}{Physical Review A {\bf 79}, 032301}~(2009).

\bibitem{Shirley1965}
Jon~H. Shirley.
\newblock ``Solution of the schr{\"{o}}dinger equation with a hamiltonian periodic in time''.
\newblock \href{https://dx.doi.org/10.1103/PhysRev.138.B979}{Physical Review {\bf 138}, B979--B987}~(1965).

\bibitem{Zeuch2020a}
Daniel Zeuch and David~P. DiVincenzo.
\newblock ``Refuting a {Proposed Axiom} for {Defining} the {Exact Rotating Wave Approximation}''~(2020).
\newblock  \href{http://arxiv.org/abs/2010.02751}{arXiv:2010.02751}.

\bibitem{Theis2017}
L.~S. Theis and F.~K. Wilhelm.
\newblock ``Nonadiabatic corrections to fast dispersive multiqubit gates involving {$Z$} control''.
\newblock \href{https://dx.doi.org/10.1103/physreva.95.022314}{Physical Review A {\bf 95}, 022314}~(2017).

\bibitem{Malekakhlagh2020a}
Moein Malekakhlagh, Alexandru Petrescu, and Hakan~E. T{\"{u}}reci.
\newblock ``Lifetime renormalization of weakly anharmonic superconducting qubits. {I}. {Role} of number nonconserving terms''.
\newblock \href{https://dx.doi.org/10.1103/PhysRevB.101.134509}{Physical Review B {\bf 101}, 134509}~(2020).

\bibitem{Petrescu2020}
Alexandru Petrescu, Moein Malekakhlagh, and Hakan~E. T{\"{u}}reci.
\newblock ``Lifetime renormalization of driven weakly anharmonic superconducting qubits. {II}. {The} readout problem''.
\newblock \href{https://dx.doi.org/10.1103/PhysRevB.101.134510}{Physical Review B {\bf 101}, 134510}~(2020).

\bibitem{Malekakhlagh2020}
Moein Malekakhlagh, Easwar Magesan, and David~C. McKay.
\newblock ``First-principles analysis of cross-resonance gate operation''.
\newblock \href{https://dx.doi.org/10.1103/PhysRevA.102.042605}{Physical Review A {\bf 102}, 042605}~(2020).

\bibitem{Shillito2021}
Ross Shillito, Jonathan~A. Gross, Agustin Di~Paolo, {{\'{E}}}lie Genois, and Alexandre Blais.
\newblock ``Fast and differentiable simulation of driven quantum systems''.
\newblock \href{https://dx.doi.org/10.1103/physrevresearch.3.033266}{Physical Review Research {\bf 3}, 033266}~(2021).

\bibitem{Puzzuoli2023}
Daniel Puzzuoli, Sophia~Fuhui Lin, Moein Malekakhlagh, Emily Pritchett, Benjamin Rosand, and Christopher~J. Wood.
\newblock ``Algorithms for perturbative analysis and simulation of quantum dynamics''.
\newblock \href{https://dx.doi.org/10.1016/j.jcp.2023.112262}{Journal of Computational Physics {\bf 489}, 112262}~(2023).

\bibitem{Marecek2020}
Jakub Marecek and Jiri Vala.
\newblock ``Quantum {Optimal Control} via {Magnus Expansion} and {Non-Commutative Polynomial Optimization}''~(2020).
\newblock  \href{http://arxiv.org/abs/2001.06464}{arXiv:2001.06464}.

\bibitem{Sanchez2011}
S.~S{\'{a}}nchez, F.~Casas, and A.~Fern{\'{a}}ndez.
\newblock ``New analytic approximations based on the {Magnus} expansion''.
\newblock \href{https://dx.doi.org/10.1007/s10910-011-9855-y}{Journal of Mathematical Chemistry {\bf 49}, 1741--1758}~(2011).

\bibitem{Butcher2009}
Eric~A. Butcher, Ma{\textquoteright}en Sari, Ed~Bueler, and Tim Carlson.
\newblock ``Magnus{\textquoteright} expansion for time-periodic systems: {Parameter}-dependent approximations''.
\newblock \href{https://dx.doi.org/10.1016/j.cnsns.2009.02.030}{Communications in Nonlinear Science and Numerical Simulation {\bf 14}, 4226--4245}~(2009).

\bibitem{Kopylov2019}
Nikita Kopylov.
\newblock ``Magnus-based geometric integrators for dynamical systems with time-dependent potentials''.
\newblock \href{https://dx.doi.org/10.4995/thesis/10251/118798}{PhD thesis}.
\newblock Universitat Politecnica de Valencia.
\newblock Valencia (Spain)~(2019).

\bibitem{Geiser2014}
Juergen Geiser and Vahid Yaghoubi.
\newblock ``Effective {Simulation Methods} for {Structures} with {Local Nonlinearity}: {Magnus} integrator and {Successive Approximations}''~(2014).
\newblock  \href{http://arxiv.org/abs/1411.7134}{arXiv:1411.7134}.

\bibitem{Bamber2020}
Jamie Bamber and Will Handley.
\newblock ``Beyond the {Runge-Kutta-Wentzel-Kramers-Brillouin} method''.
\newblock \href{https://dx.doi.org/10.1103/PhysRevD.101.043517}{Physical Review D {\bf 101}, 043517}~(2020).

\bibitem{Arnal2021}
Ana Arnal, Fernando Casas, Cristina Chiralt, and Jos{\'{e}}~Angel Oteo.
\newblock ``A {Unifying Framework} for {Perturbative Exponential Factorizations}''.
\newblock \href{https://dx.doi.org/10.3390/math9060637}{Mathematics {\bf 9}, 637}~(2021).

\bibitem{Vogl2019}
Michael Vogl, Pontus Laurell, Aaron~D. Barr, and Gregory~A. Fiete.
\newblock ``Analog of {Hamilton-Jacobi} theory for the time-evolution operator''.
\newblock \href{https://dx.doi.org/10.1103/PhysRevA.100.012132}{Physical Review A {\bf 100}, 012132}~(2019).

\bibitem{Dyson1949}
F.~J. Dyson.
\newblock ``The {${S}$} {Matrix} in {Quantum Electrodynamics}''.
\newblock \href{https://dx.doi.org/10.1103/PhysRev.75.1736}{Physical Review {\bf 75}, 1736--1755}~(1949).

\bibitem{Li2023}
Boxi Li, Tommaso Calarco, and Felix Motzoi.
\newblock ``Experimental error suppression in cross-resonance gates via multi-derivative pulse shaping''.
\newblock \href{https://dx.doi.org/10.48550/ARXIV.2303.01427}{npj Quantum Inf 10, 66 (2024){\bf 10}}~(2023).
\newblock  \href{http://arxiv.org/abs/2303.01427}{arxiv:2303.01427}.

\bibitem{Moskalenko2021}
I.~N. Moskalenko, I.~S. Besedin, I.~A. Simakov, and A.~V. Ustinov.
\newblock ``Tunable coupling scheme for implementing two-qubit gates on fluxonium qubits''.
\newblock \href{https://dx.doi.org/10.1063/5.0064800}{Applied Physics Letters {\bf 119}, 194001}~(2021).
\newblock  \href{http://arxiv.org/abs/2107.11550}{arxiv:2107.11550}.

\bibitem{Weiss2022}
D.K. Weiss, Helin Zhang, Chunyang Ding, Yuwei Ma, David~I. Schuster, and Jens Koch.
\newblock ``Fast {{High-Fidelity Gates}} for {{Galvanically-Coupled Fluxonium Qubits Using Strong Flux Modulation}}''.
\newblock \href{https://dx.doi.org/10.1103/PRXQuantum.3.040336}{PRX Quantum {\bf 3}, 040336}~(2022).

\bibitem{Ding2023}
Leon Ding, Max Hays, Youngkyu Sung, Bharath Kannan, Junyoung An, Agustin Di~Paolo, Amir~H. Karamlou, Thomas~M. Hazard, Kate Azar, David~K. Kim, Bethany~M. Niedzielski, Alexander Melville, Mollie~E. Schwartz, Jonilyn~L. Yoder, Terry~P. Orlando, Simon Gustavsson, Jeffrey~A. Grover, Kyle Serniak, and William~D. Oliver.
\newblock ``High-{Fidelity}, {Frequency-Flexible Two-Qubit Fluxonium Gates} with a {Transmon Coupler}''.
\newblock \href{https://dx.doi.org/10.1103/PhysRevX.13.031035}{Physical Review X {\bf 13}, 031035}~(2023).

\bibitem{Rosenfeld2024}
Emma~L. Rosenfeld, Connor~T. Hann, David~I. Schuster, Matthew~H. Matheny, and Aashish~A. Clerk.
\newblock ``Designing high-fidelity two-qubit gates between fluxonium qubits''.
\newblock \href{https://dx.doi.org/10.1103/prxquantum.5.040317}{PRX Quantum {\bf 5}, 040317}~(2024).
\newblock  \href{http://arxiv.org/abs/2403.07242}{arXiv:2403.07242}.

\end{thebibliography}

\onecolumn\newpage
\appendix

\section{\texorpdfstring{qCH\textsubscript{eff}}{qCHeff} Package Details} \label{sec:package-appendix}

\subsection{Package Installation}

qCH\textsubscript{eff} is hosted on PyPI (\url{https://pypi.org/project/qcheff/}) and can be installed in a Python virtual environment using pip with the command \texttt{pip install qcheff}. The same command will also work inside Conda environments, provided pip is available. 

\subsection{Online Documentation}

Online documentation for qCH\textsubscript{eff} is hosted online via Read the Docs (\url{https://qcheff.readthedocs.io/en/latest/}), including detailed installation instructions, examples and API documentation. The documentation is also available in the Python interpreter using the \texttt{help()} function. 

\subsection{\texttt{qcheff} Operator interface}
In the interest of minimal code duplication, we have defined a convenience wrapper class around NumPy/SciPy(sparse)/CuPy arrays. This provides a uniform interface to access some commonly used operations on these matrices. For iterative SWT methods, it is necessary to separate the diagonal and off-diagonal values in a matrix since they correspond to energies and couplings respectively, so \texttt{qcheff.operator} provides a uniform interface to these operations that dispatch to the appropriate underlying array module. This wrapper is for convenience of the user only. The NPAD subroutines work directly with the underlying array.

\subsection{Iterative Schrieffer-Wolff Transformations}
This submodule contains the NPAD technique. Although this is the only method at the time of writing, other variants of iterative SWT may be added in the future. NPAD is an exact SWT at the level of couplings, therefore, this technique will iterate over couplings. We provide a general \texttt{ExactIterativeSWT} class interface. This interface provides the following methods:
\begin{itemize}
    \item \texttt{givens\_rotation\_matrix()}: Create the Given's rotation matrix. 
    \item \texttt{unitary\_transformation()}: $UHU^{\dagger}$ for NPAD. May be generalized to time-dependent case.
    \item \texttt{eliminate\_coupling()}: Eliminate a single coupling.
    \item \texttt{eliminate\_couplings()}: Eliminate multiple couplings simultaneously.
    \item \texttt{largest\_couplings()}: Obtain the largest couplings.
\end{itemize}
There are two subclasses implementing NPAD: \texttt{NPADScipySparse} and \texttt{NPADCupySparse},  owing to differences between the SciPy and CuPy sparse matrix/array API. They each implement a \texttt{givens\_rotation\_matrix()} method using the appropriate sparse matrix module, since these matrices are largely sparse ($N+2$ nonzero elements for $N\times N$ matrices). For the user, an \texttt{NPAD(H)} function is provided for convenience. This will return an object of the appropriate subclass based on the input array \texttt{H}.

\subsection{Magnus time-evolution}

The \texttt{MagnusTimeEvol} class provides a standard interface for all Magnus time-evolution operations. The interface includes two main functions:  

\begin{itemize}
    \item \texttt{update\_control\_sigs()}: Change the control pulse applied to the system. 
    \item \texttt{evolve()}: Evolve the system for a given number of Magnus intervals.
\end{itemize}

There are two subclasses that implement this interface: \texttt{MagnusTimeEvolDense} and \texttt{MagnusTimeEvolSparse}. A schematic overview of \texttt{MagnusTimeEvolDense} is shown in Fig.~\ref{fig:qcheff-schematic}. Notably, it computes the propagators for all Magnus intervals simultaneously. Sparse arrays are limited to be 2-d in SciPy/CuPy, so we cannot use the same technique without converting to dense matrices. Furthermore, exponentiating a matrix uses a lot of memory, so the dense implementation is limited in the size of systems it can handle. Instead, we loop over the propagators for each Magnus interval. This lets the user examine larger systems at the expense of slow iteration in Python loop. At the time of writing, CuPy does not have an implementation of matrix exponentiation \texttt{expm()} that works with 3-d arrays, so we have implemented such a routine based on the Taylor expansion, inspired by \url{https://github.com/lezcano/expm} for use on the GPU.

For the user, an \texttt{magnus(tlist, drift\_ham, control\_sigs, control\_hams)} function is provided for convenience. This will return an object of the appropriate subclass based on the input array.

\section{Benchmarks}\label{sec:benchmarks}
All benchmarks were performed in the cloud using an NVIDIA Brev container of instance type \texttt{gpu\_1x\_h100\_sxm5}. This instance was hosted by Lambda Labs and contained single NVIDIA H100 GPU system 80 GB VRAM, 26 vCPU cores, 225 GiB system memory (RAM) and a 2.75 TiB SSD for storage. The container was further configured with CUDA 12.2.2, and Python 3.12. Both CPU and GPU benchmarks were run in the same Python environment, orchestrated using Pixi. The locked environment specification can be found in the Pixi lockfile in the repository under \texttt{examples/pixi.lock}. For CPU benchmarks, SciPy and NumPy can use multiple CPU cores. No restrictions were placed on the number of CPU cores used by these libraries. The benchmarks were run in a Pixi environment with qCheff version \texttt{v0.2.1} on PyPI and GitHub. All reported times exclude setup or tear-down such as creating the input matrices, Hamiltonians, etc. Any CPU $\leftrightarrow$ GPU transfers occur at this stage to ensure that the input arrays are located on the correct device memory (CPU/GPU) before the corresponding benchmark is started, thereby profiling the computational steps only. All benchmarking code is available inside the \texttt{qCHeff/examples/paper/} directory after cloning the GitHub repository at the release tagged \texttt{v0.2.1}.
\end{document}